\documentclass[
 reprint,
 aps,
 prl,
 superscriptaddress,
 amsmath,
 amssymb,
 floatfix
]{revtex4-2}

\usepackage{bm}
\usepackage{graphicx}
\usepackage[small,bf,format=plain,singlelinecheck=false,justification=Justified]{caption}
\usepackage{physics}
\usepackage{siunitx}
\usepackage{hyperref}
\usepackage{xcolor}
\usepackage[symbol]{footmisc}
\hypersetup{
    colorlinks=true,
    citecolor=blue,
    linkcolor=blue,
    urlcolor=blue,
}

\makeatletter
\renewcommand*{\@fnsymbol}[1]{\ensuremath{\ifcase#1\or *\or *,\dagger\or \ddagger\or
   \mathsection\or \mathparagraph\or \|\or **\or \dagger\dagger
   \or \ddagger\ddagger \else\@ctrerr\fi}}
\makeatother

\begin{document}

    \preprint{APS/123-QED}

    \title{Contrasting magnetothermal conductivity \\in sibling Co-based honeycomb-lattice antiferromagnets}

    \author{Masato Ueno}
    \thanks{These two authors equally contributed to this work.}
    \affiliation{Department of Advanced Materials Science, University of Tokyo, Kashiwa 277-8561, Japan}

    \author{Takashi Kurumaji} 
    \thanks{Present address: Division of Physics, Mathematics and Astronomy, California Institute of Technology, Pasadena, California 91125, USA}
    \affiliation{Department of Advanced Materials Science, University of Tokyo, Kashiwa 277-8561, Japan}
    
    \author{Shunsuke Kitou}
    \affiliation{Department of Advanced Materials Science, University of Tokyo, Kashiwa 277-8561, Japan}

    \author{\\Masaki Gen}
    \affiliation{RIKEN Center for Emergent Matter Science (CEMS), Wako, 351-0198, Japan}

    \author{Yuiga Nakamura}
    \affiliation{Japan Synchrotron Radiation Research Institute (JASRI), Sayo-gun, 679-5198, Japan}

    \author{Yusuke Tokunaga}
    \affiliation{Department of Advanced Materials Science, University of Tokyo, Kashiwa 277-8561, Japan}

    \author{Taka-hisa Arima}
    \affiliation{Department of Advanced Materials Science, University of Tokyo, Kashiwa 277-8561, Japan}
    \affiliation{RIKEN Center for Emergent Matter Science (CEMS), Wako, 351-0198, Japan}

    \date{\today}
    
    \begin{abstract}
        Honeycomb-lattice antiferromagnets have attracted wide attention for exploration of exotic heat transport and their interplay with magnetic excitations. 
        In this work, we have revealed a contrasting behavior in the magneto-thermal conductivity (MTC) between two Co-based honeycomb-lattice magnets $\mathrm{Co}_4\mathcal{M}_2\mathrm{O}_9$ ($\mathcal{M} = \mathrm{Nb},\,\mathrm{Ta}$),
        despite their identical lattice structures and quite similar magnetism. 
        $\mathrm{Co_4Ta_2O_9}$ exhibits enhanced MTC of about \SI{550}{\%} at \SI{9}{T} of an in-plane magnetic field, comparable to other honeycomb-magnets, while MTC for $\mathrm{Co_4Nb_2O_9}$ reaches only $\sim \SI{30}{\%}$.
        This marked difference is ascribed to distinct features in the field-induced evolution of magnetic excitations that resonantly scatter phonons.
        This finding sheds light on implicit impacts of non-magnetic ions on thermal transport, 
        and hints at the potential for broad heat-transport tunability while preserving magnetism and lattice structures.
    \end{abstract}
    
    \maketitle

    Thermal transport properties are attracting central interest as a useful probe to unravel intriguing nature of quasi-particles in quantum materials \cite{Li_probe_2020}. 
    Magnetic insulators are extensively investigated to explore unconventional heat transport phenomena, 
    including ballistic spin conduction \cite{Sologubenko_excitation_2007,Hess_heat_2007}, thermal Hall effect \cite{Strohm_Tb3Ga5O12_2005,onose_observation_2010,grissonnanche_giant_2019}, and large magnetothermal conductivity (MTC) \cite{Ando_CuGeO3_1998,Hofmann_SrCu2(BO3)2_2001,sales_K2V3O8_2002,Chen_(CH3)2NH2CuCl3_2011,Ke_Ba3Mn2O8_2011,wu_Cu3(CO3)2(OH)2_2016,zhao_Co3V2O8_2016}. 
    Phonons are in most cases dominant heat carriers, while spin-phonon coupling often plays a pivotal role in originating fascinating features.

    Recent efforts have been primarily concentrated on honeycomb-lattice magnets. 
    This distinct class of materials is anticipated to harbor exotic characteristics, 
    including topological magnon bands \cite{Owerre_theoretical_2016,Owerre_honeycomb_2016,Fransson_dirac_2016,Kim_HKMmodel_2016,Chen_CrI3_2018,Yuan_CoTiO3_2020,Fengfeng_CrSiTe3CrGeTe3_2021}, 
    Majorana edge mode in the context of Kitaev spin liquids \cite{kitaev_anyons_2006,kasahara_RuCl3_2018}, 
    and chiral phonons carrying angular momenta \cite{Zhang_chiralphonon_2015,Hanyu_chiralphonon_2018,Wu_chiralphonon_2023}. 
    Significant thermal transport properties have been observed, including large MTC due to strong spin-phonon coupling \cite{Leahy_RuCl3_2017,Pocs_CrCl3_2020,zhong_Ba2Co(AsO4)2_2020,hong_Na2Co2TeO6_2021,tu_Ba2Co(AsO4)2_2023}, 
    and thermal Hall effect \cite{ideue_Fe2Mo3O8_2017,Kasahara_RuCl3Hall_2018,kasahara_RuCl3_2018,zhang_VI3_2021,yang_Ni3TeO6_2022,Choi_Cr2Ge2Te6_2023}. 
    It has recently been recognized that exotic heat conduction can occur also when a field is parallel to the honeycomb-lattice plane \cite{Chern_sign_2021,Utermohlen_tensor_2021,Yokoi_RuCl3_2021,takeda_Na2Co2TeO6_2022,Czajka_RuCl3_2023,Kurumaji_symmetry_2023,chen_Na2Co2TeO6_2023}.
    The intricate interplay between phonons and magnetic excitations continues to be a fertile ground for ongoing discussion and exploration \cite{Hentrich_RuCl3_2018, kasahara_RuCl3_2018, Hentrich_RuCl3_2019, hong_Na2Co2TeO6_2021, lefrancois_evidence_2022, Czajka_RuCl3_2023}. 
    The pursuit of understanding the role of spin-phonon coupling promises a potential pathway to gain tunability of heat conduction within this intriguing group of magnets.

\begin{figure}[t]
        \centering
        \includegraphics[width=246pt]{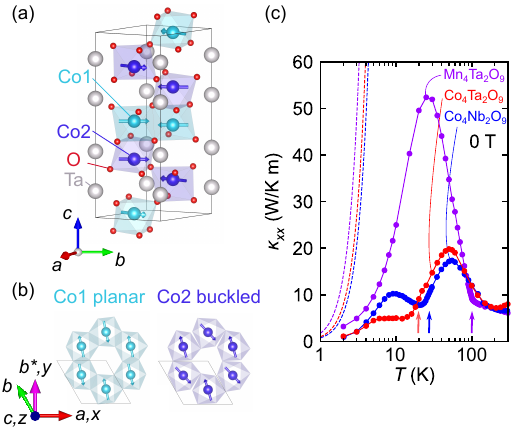}
        \caption{
            (a) Schematic crystal and magnetic structure of CTO (space group: $P\bar{3}c1$) visualized by VESTA software \cite{momma_vesta_2011}.
            Cyan (blue) octahedron represents oxygen coordination around Co1 (Co2). 
            An arrow on each Co site represents the magnetic moments. 
            (b) Top view of a planar (buckled) honeycomb lattice formed by Co1 (Co2). Cartesian coordinates, $x$, $y$, $z$ are defined.
            (c) Temperature dependence of zero-field thermal conductivity ($\kappa_{xx}$) of CTO with the heat current along the $x$ axis ($j_Q \parallel x \parallel a$).
            The data for CNO and MTO is also shown. 
            Arrows denote the transition temperature ($T_{\mathrm{N}}^{\mathrm{CTO}} \sim \SI{20}{K}$, $T_{\mathrm{N}}^{\mathrm{CNO}} \sim \SI{27}{K}$, $T_{\mathrm{N}}^{\mathrm{MTO}} \sim \SI{100}{K}$), identified by the magnetization measurements with $\mu_{0}H = \SI{0.1}{T}$.
            Dashed lines are the phonon thermal conductivity in the Casimir limit.
        }
        \label{crystal_structures}
    \end{figure}

    Corundum-derivative compounds, $\mathrm{Co_4Ta_2O_9}$ (CTO) and $\mathrm{Co_4Nb_2O_9}$ (CNO), have attracted recent interests as a host of stacked honeycomb-lattice of magnetic $\mathrm{Co^{2+}}$ ions \cite{Motome_kitaev_2020}.
    They belong to a centrosymmetric space group $P\bar{3}c1$ \cite{Bertaut_A2B4O9_1961} (Fig.\ \ref{crystal_structures}(a)), 
    and the $\mathcal{PT}$-symmetric antiferromagnetic order of $\mathrm{Co}^{2+}$ ions breaks both the space inversion ($\mathcal{P}$) and time-reversal ($\mathcal{T}$) symmetries below N\'{e}el temperatures ($T_{\mathrm{N}}$) giving rise to magnetoelectric coupling \cite{Fischer_A2B4O9_1972,Kolodiazhnyi_Co4Nb2O9_2011,Fang_Co4Nb2O9_2014,fang_Co4Ta2O9_2015,khanh_Co4Nb2O9_2016,Xie_Co4Nb2O9_2016,Yin_Co4Nb2O9_2016,Lu_Co4Nb2O9_2016,khanh_manipulation_2017,cao_Co4Nb2O9_2017,chaudhary_nature_2019,lee_highly_2020}. 
    $\mathrm{Ta}^{5+}$ and $\mathrm{Nb}^{5+}$ ions are not magnetic, hosting no d-electrons.
    There are two types of $\mathrm{CoO_6}$ octahedra (Co1 and Co2) in CTO and CNO. 
    Heavily buckled honeycomb networks of Co2 and less buckled ones of Co1 (Fig.\ \ref{crystal_structures}(b)) are alternately stacked along the $c$ axis.
    They are termed buckled and planar, respectively.

    Previous neutron diffraction experiments revealed slightly canted antiferromagnetic order at zero field below $T_{\mathrm{N}} = \SI{20}{K}$ for CTO \cite{chaudhary_nature_2019,choi_noncollinear_2020}, and below $T_{\mathrm{N}} = \SI{27}{K}$ for CNO \cite{khanh_Co4Nb2O9_2016,deng_spin_2018,ding_Co4Nb2O9_2020}. 
    According to Refs. \cite{choi_noncollinear_2020,ding_Co4Nb2O9_2020}, for both compounds, magnetic moments are confined in the $ab$-plane because of the easy-plane type magnetic anisotropy of the $\mathrm{Co}^{2+}$ ions. 
    Magnetic moments are mutually canted within the $ab$-plane due to the Dzyaloshinskii-Moriya interaction along the $c$-axis as shown in Fig.\ \ref{crystal_structures}(b) for CTO.
    The magnetic structure of CNO is similar to CTO, while the magnetic moments are in a different orientation in the $ab$-plane from those in CTO.
    An in-plane magnetic field induces a spin-flop-like transition at low magnetic fields ($H_c = \SI{0.3}{T}$ for CTO \cite{lee_highly_2020}, $= \SI{0.2}{T}$-$\SI{0.75}{T}$ for CNO \cite{Cao_Co4Nb2O9_2015,khanh_Co4Nb2O9_2016}). 
    For CNO, it has been revealed that this phenomenon is ascribed to the antiferromagnetic domain rearrangement so that the N\'{e}el-vector becomes perpendicular to the applied magnetic field \cite{ding_Co4Nb2O9_2020}. 

    In this letter, we report a distinctive behavior of MTC in isostructural honeycomb-antiferromagnets, CTO and CNO \cite{Bertaut_A2B4O9_1961}. 
    Despite the structural and magnetic similarity, the MTC in CTO reaches \SI{550}{\%} in an in-plane magnetic field of \SI{9}{T}, 
    comparable to other large MTC honeycomb-lattice magnets, while CNO shows the MTC suppressed by more than one order of magnitude in the same field configuration. 
    This contrasting MTC behaviors are attributed to the subtle balance between the energy scales of magnetic excitations and features of phonon resonant scattering. 
    This result provides an insight on the tunability of heat transport in magnets without significantly changing lattice and magnetism.

    Single crystals of CTO, CNO, and $\mathrm{Mn_4Ta_2O_9}$ (MTO) for measurements were grown by the floating zone method (see Sec.\ A in Ref.\ \cite{Ueno_SupplemntalMaterial_2023}).
    Thermal conductivity at low temperatures in a magnetic field was measured by the standard steady-state method using a commercial cryostat equipped with a superconducting magnet. 
    Cartesian coordinates, $x$, $y$, $z$ are defined as shown in Fig.\ \ref{crystal_structures}(b), and in-plane thermal conductivity $\kappa_{ii}$ ($i = x,\, y$) is estimated following the formula $j_{Qi} = \kappa_{ii}(-\partial_i T)$, where $j_{Qi}$ and $\partial_i T$ are heat current density and longitudinal temperature gradient along $i$ axis, respectively.
    The $j_{Qi}$ was generated from a 1-$\mathrm{k\Omega}$ heater chip attached to an edge of a rectangular-shaped sample, where a DC electric current was applied by Keithley 6221 through manganin wires ($\phi=\SI{25}{\mu m}$).
    The other edge was attached to the copper block as a heat bath. 
    The $\partial_i T$ on the sample was monitored by type-E thermocouples ($\phi=\SI{25}{\mu m}$). 
    The temperature of the heat bath was measured by a resistive thermometer (Cernox CX-1030-SD), which was calibrated at low temperatures in magnetic fields. 
    High-vacuum condition ($<\SI{1e-3}{Pa}$) was maintained during the measurement to prevent convection. 
    Radiation from the environment was suppressed by using a brass cap shield on the sample puck.
    DC magnetization was measured by a superconducting quantum interferometer device magnetometer system (MPMS, Quantum Design).
    Thermal expansion and magnetostriction along $a$ and $b^*$ were simultaneously measured by the fiber-Bragg-grating method using a cryostat equipped with a superconducting magnet (Spectromag, Oxford Instruments)\cite{Gen_FBG_2022}.

    \begin{figure}[t]
        \centering
        \includegraphics[width=246pt]{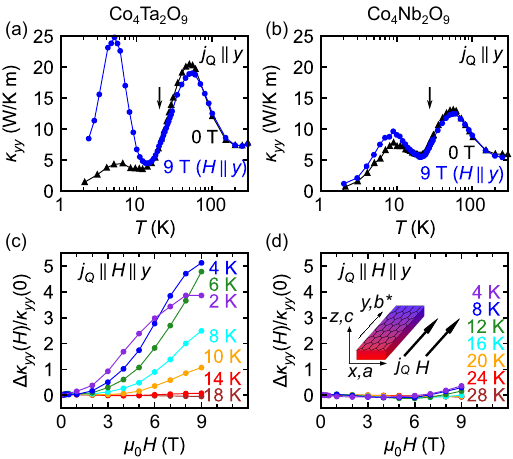}
        \caption{
            The temperature dependence of thermal conductivity along $y$ axis ($\kappa_{yy}$) in (a) CTO and (b) CNO at zero field and under the magnetic field of \SI{9}{T} along $H\parallel y$.
            The transition temperature ($T_{\mathrm{N}}$) at $\mu_0 H = \SI{0.1}{T}$ is also denoted by an arrow. 
            Field dependence of the magnetothermal conductivity $\Delta \kappa_{yy}(H)/\kappa_{yy}(0)$ of (c) CTO and (d) CNO at various temperatures.
            The definition of the Cartesian coordinates is denoted as an inset of (d). 
        }
        \label{kappa_vs_TH_y_CTOCNO}
    \end{figure}

    Figure \ref{crystal_structures}(c) shows the temperature dependence of thermal conductivity of CTO and CNO in zero magnetic field.
    We also measured the thermal conductivity in isostructural MTO to compare the effect of spin-orbital coupled magnetic moment in Co$^{2+}$ ions and a spherical $S=5/2$ moment in Mn$^{2+}$ ions.
    $\kappa_{xx}$ in MTO shows a typical feature dominated by a phononic contribution, which is characterized by a peak at \SI{25}{K}, as observed in various $\mathrm{Mn^{2+}}$ compounds \cite{Slack_Mn_1961, Ozhogin_CoCO3MnCO3_1983,rives1969effect}, while a critical scattering induces a kink at $T_{\mathrm{N}}$ (see a purple arrow).
    CTO and CNO, on the other hand, show a prominent suppression of thermal conductivity with a dip at $T$ lower than $T_{\mathrm{N}}$. 
    This indicates that magnetic excitations associated with Co-3d electrons strongly scatter phonons through spin-lattice coupling \cite{Ozhogin_CoCO3MnCO3_1983,zhao_Co3V2O8_2016}.
    We also estimate the boundary-scattering (Casimir) limit of pure phonon thermal conductivity (see Sec. \ E in Ref. \ \cite{Ueno_SupplemntalMaterial_2023}) as shown by dashed lines in Fig. \ref{crystal_structures}(c).
    The observed $\kappa_{xx}$ in MTO is moderately suppressed from the ideal value due to relatively small spin-phonon interaction \cite{gustafson1973thermal,sanders_effect_1977}, while those in CTO and CNO are more strikingly influenced by the magnetic scattering.

    The similarity in thermal conductivity between CTO and CNO may originate from the similarity in the crystal structure (see Sec.\ B in Ref.\ \cite{Ueno_SupplemntalMaterial_2023}), electron configuration ($\mathrm{Co^{2+}}$ is $3\mathrm{d}^7$, $\mathrm{Ta^{5+}}$ and $\mathrm{Nb^{5+}}$ are $\mathrm{d}^0$), and magnetic structure \cite{ding_Co4Nb2O9_2020,choi_noncollinear_2020}.
    Both compounds show the easy-plane type magnetic anisotropy due to the orbital moment in $\mathrm{Co^{2+}}$ ions, while CTO has a softer magnetization process in the in-plane magnetic field than CNO does (see Sec.\ C in Ref.\ \cite{Ueno_SupplemntalMaterial_2023}). 
    Thermal expansion and magnetostriction below $T_{\mathrm{N}}$ (see Sec.\ C in Ref.\ \cite{Ueno_SupplemntalMaterial_2023}) are also comparable, suggesting that the exchange striction plays a similar role in these two compounds.
    
    Contrasting behavior in the thermal conductivity between CTO and CNO emerges in a magnetic field applied parallel to the honeycomb planes ($H\perp c$). 
    Figure \ref{kappa_vs_TH_y_CTOCNO}(a) shows the temperature dependence of thermal conductivity $\kappa_{yy}$ of CTO when the magnetic field is applied parallel to the heat flow ($j_Q\parallel H \parallel y$).
    The temperature dependence of $\kappa_{yy}$ shows a double-peak structure, at about \SI{50}{K} and about \SI{5}{K}.
    At \SI{9}{T}, the high-temperature peak is slightly suppressed, while the low-temperature one exhibits a large enhancement, suggesting a suppression of phonon scattering. 
    In CNO, on the other hand, the field-induced changes for both peaks are much smaller than that of CTO (Fig.\ \ref{kappa_vs_TH_y_CTOCNO}(b)).
    
    This difference becomes clearer in the plot of MTC ratio $\Delta \kappa_{yy}(H)/\kappa_{yy}(0) = [\kappa_{yy}(H)-\kappa_{yy}(0)]/\kappa_{yy}(0)$ of CTO and CNO at various temperatures, as shown in Figs.\ \ref{kappa_vs_TH_y_CTOCNO}(c)-(d).
    In CTO, the field-induced enhancement of $\kappa_{yy}$ is observed below \SI{10}{K}, and the MTC ratio reaches \SI{550}{\%} at $T=\SI{5}{K}$. 
    That of CNO is $\sim \SI{30}{\%}$ at the maximum at $T=\SI{9}{K}$.
    These features indicate that low energy magnetic excitations give rise to the striking difference in MTC. 
    We also note that the spin-flop transition at low fields only slightly affects the MTC for both compounds (see Sec.\ C in Ref.\ \cite{Ueno_SupplemntalMaterial_2023}), suggesting that the difference in the tilt angle of the magnetic moment between CTO and CNO plays a minor role in the heat transport.
    
    We have also measured the anisotropy of MTC with respect to the heat current direction and the magnetic field (see Sec.\ D in Ref.\ \cite{Ueno_SupplemntalMaterial_2023}). 
    The MTC in the $j_{Q}\parallel H\parallel x$ condition is comparable to that of $j_{Q}\parallel H\parallel y$, as expected from small in-plane magnetic anisotropy in a trigonal system.
    In contrast to the in-plane-field MTC, that in the out-of-plane field condition ($j_{Q} \perp z$, $H \parallel z$) does not provide a large MTC ratio for either CTO or CNO. 
    This is consistent with the easy-plane type magnetic anisotropy, where the magnetic excitations are expected to weakly respond to $H \parallel z$.

    \begin{figure}[t]
        \centering
        \includegraphics[width=\linewidth]{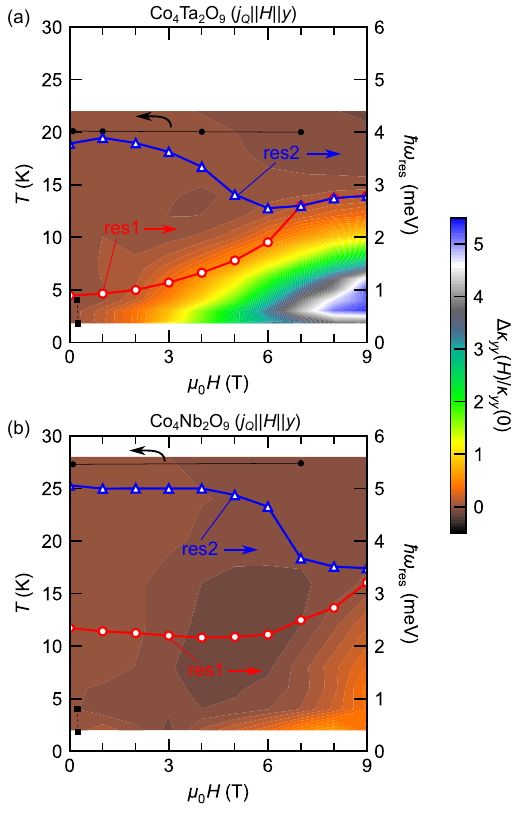}
        \caption{
            Color map of the $\Delta \kappa_{yy}(H)/\kappa_{yy}(0)$ in the $H$-$T$ phase diagram of (a) CTO and (b) CNO for $H\parallel y$.
            Black circles (solid lines) are the magnetic transition temperatures obtained by magnetization measurements. 
            Black squares (dashed lines) are the spin-flop transition fields (see Fig.\ S3 in Ref.\ \cite{Ueno_SupplemntalMaterial_2023}).
            The extracted resonant frequencies ($\omega_{\mathrm{res}1}$, $\omega_{\mathrm{res}2}$) as a function of applied magnetic field are also shown (red circles and blue triangles).
            The difference between $\omega_{\mathrm{res}1}$ and $\omega_{\mathrm{res}2}$ in CTO at higher magnetic fields is too small to resolve.
        }
        \label{colormap_y}
    \end{figure}

    To capture the interplay between phonons and magnetic excitations, the thermal conductivity data are analyzed by using the Callaway model \cite{Callaway_CallawayModel_1959,Callaway_Callaway_1961}.
    In this model, the temperature dependence of the thermal conductivity is calculated numerically by considering the frequency dependence of the phonon relaxation time $\tau_{\mathrm{tot}}$, as
    \begin{align}
        \kappa(T) = \frac{k_{\mathrm{B}}}{2\pi^2 v_{s}} \left(\frac{k_{\mathrm{B}}T}{\hbar}\right)^3 \int_{0}^{\frac{\Theta_{\mathrm{D}}}{T}} F(\omega,\,T) \tau_{\mathrm{tot}} (\omega,\,T)d\xi,
        \label{formula:001}
    \end{align}
    where $k_{\mathrm{B}}$ is Boltzmann constant, $\hbar$ is Dirac constant, $\Theta_{\mathrm{D}}$ is Debye temperature, $v_s$ is the acoustic phonon velocity (see Sec.\ E in Ref.\ \cite{Ueno_SupplemntalMaterial_2023}), $\omega$ is phonon frequency, $\xi=\hbar\omega/k_{\mathrm{B}}T$, and $F(\omega,\,T)=(\xi^4 e^\xi)/(e^\xi-1)^2$ \cite{Callaway_CallawayModel_1959,Callaway_Callaway_1961}.
    $\tau_{\mathrm{tot}}^{-1} = \tau_{\mathrm{c}}^{-1} + \tau_{\mathrm{mag}}^{-1}$, 
    where the nonmagnetic term $\tau_{\mathrm{c}}$ commonly used empirically as the pure phononic relaxation term including boundary, point defect, dislocation, and umklapp scattering.
    We assume that the $\tau_{\mathrm{mag}}$ that represents phonon resonant scattering by magnetic excitations at typical frequencies $\omega_{\mathrm{res}j}$ ($j = 1, 2, \ldots$), phenomenologically given by \cite{SHEARD_phonon_1973,Wybourne_phonon_1985,Prasai_AB2O6_2018,Hentrich_RuCl3_2018,Hentrich_RuCl3_2020,yang_Cr2Si2Te6_2023}
    \begin{align}
        \tau_{\mathrm{mag}}^{-1} &= \sum_{j} C_j \frac{\omega^4}{(\omega^2 - \omega_{\mathrm{res}j}^2)^2} \frac{\exp\left(-\frac{\hbar\omega_{\mathrm{res}j}}{k_{\mathrm{B}}T}\right) }{1 + \exp\left(-\frac{\hbar\omega_{\mathrm{res}j}}{k_{\mathrm{B}}T}\right) },
        \label{formula_omega}
    \end{align}
    where $C_j$ denotes phonon-magnetic-excitation coupling constant.

    The experimental data for both CTO and CNO are reproduced by using the close order of magnitude values for the corresponding parameters (see Sec.\ E in Ref.\ \cite{Ueno_SupplemntalMaterial_2023}), confirming the similarity of non-magnetic phonon scattering processes and spin-phonon coupling \cite{park_Co4Nb2O9-Co4Ta2O9_2023} between these compounds. 
    The major difference that gives rise to the distinct MTC is the energy of $\omega_{\mathrm{res}j}$ and their field-evolution. 
    Figures \ref{colormap_y}(a) and (b) show a color map of $\Delta \kappa_{yy}(H)/\kappa_{yy}(0)$ and field-dependence of $\hbar\omega_{\mathrm{res}1}$ and $\hbar\omega_{\mathrm{res}2}$.
    We find that at least two resonant frequencies, $\omega_{\mathrm{res}1}$ and $\omega_{\mathrm{res}2}$, have to be introduced to obtain reasonable temperature dependence of $\kappa$ in magnetic fields.
    In CTO (Fig.\ \ref{colormap_y}(a)), the resonant frequency $\omega_{\mathrm{res}1}$ ($\sim \SI{1}{meV}$) suppresses the thermal conductivity below $T_{\mathrm{N}}$ at zero magnetic field.
    As the magnetic field increases, $\omega_{\mathrm{res}1}$ is increased by the Zeeman effect to approach the higher resonant frequency $\omega_{\mathrm{res}2}$. 
    Consequently, the magnetic scattering of phonons is suppressed, resulting in the enhancement of the low-temperature peak in $\kappa_{yy}$ by a magnetic field.
    The smaller MTC in CNO is attributed to slightly higher energy of $\hbar\omega_{\mathrm{res}1}$ and the negligible sensitivity to the field at the low field regions ($<\SI{6}{T}$).

    The phonon resonant scattering frequency is considered to correspond to the energy at which the gap of magnetic excitation bands and phonon dispersions intersect.
    In the previous terahertz spectroscopy, CNO was observed to possess multiple magnetic excitations comparable to the energy of $\omega_{\mathrm{res}1}$ and $\omega_{\mathrm{res}2}$ \cite{dagar_Co4Nb2O9_2022}. 
    Some of the modes were also confirmed by inelastic neutron scattering \cite{deng_spin_2018}. 
    Although the magnetic excitations spectra in CTO have remained elusive, excitation energy $\omega_{\mathrm{res}1}$ lower than that in CNO is possibly attributed to the systematically weaker magnetic exchange interactions in CTO than those in CNO \cite{solovyev_origin_2016} as magnon gap energy in an antiferromagnet at zero field is correlated to the exchange field.
    The faster field-induced evolution of $\omega_{\mathrm{res}1}$ in CTO is consistent with softer magnetization process (see Fig.\ S2(b) and (f) in Ref.\ \cite{Ueno_SupplemntalMaterial_2023}), which suggests that the magnons (and thus, anticrossing points with phonons) gain more Zeeman energy than that in CNO.
    We note that the $\omega_{\mathrm{res}1}$ in CNO shows a slight decrease ($\sim$ 5\%) at low fields (Fig. \ref{colormap_y}(b)). This might not be an intrinsic effect but within the uncertainty of the fit parameters (see Sec.\ E in Ref.\ \cite{Ueno_SupplemntalMaterial_2023}).
    According to the DFT calculations in Ref. \cite{solovyev_origin_2016}, the softness of magnetization originates from narrower Co-3d bandwidth, larger on-site Coulomb interaction $U$, and smaller band gap between the minority-spin states. 
    These features are due to extended Ta-5d states hybridize more strongly with O-2p states. 
    This exemplifies an impact of nonmagnetic ions to an apparent identical magnets for distinct behavior in magnetothermal transport properties.
    
    \begin{figure}[t]
        \centering
        \includegraphics[width=\linewidth]{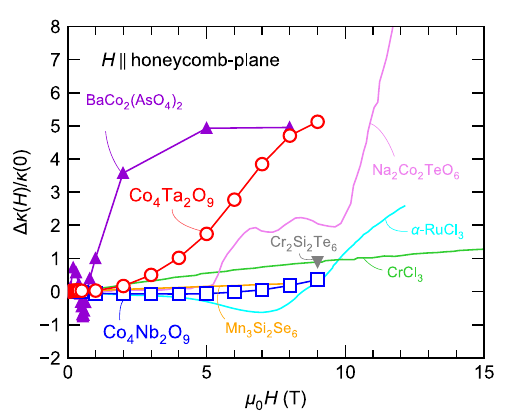}
        \caption{
            In-plane longitudinal MTC ratio in CTO and CNO obtained in this study at $T=\SI{4}{K}$ and that of a variety of honeycomb-lattice magnets. 
            $\mathrm{Na_2Co_2TeO_6}$ ($T=\SI{7}{K}$, $j_Q\parallel H$) \cite{hong_Na2Co2TeO6_2021},
            $\alpha\mathchar`-\mathrm{RuCl_3}$ ($T=\SI{2.5}{K}$, $j_Q\parallel H$) \cite{Leahy_RuCl3_2017},
            $\mathrm{CrCl_3}$ ($T=\SI{21}{K}$, $j_Q\parallel H$) \cite{Pocs_CrCl3_2020},
            $\mathrm{Mn_3Si_2Se_6}$ ($T=\SI{20}{K}$, $j_Q\parallel H$) \cite{May_Mn3Si2Se6_2020} and
            $\mathrm{Cr_2Si_2Te_6}$ ($T=\SI{32}{K}$, $j_Q\parallel H$) \cite{yang_Cr2Si2Te6_2023}.
            The transverse MTC in 
            $\mathrm{BaCo_2(AsO_4)_2}$ at \SI{0.7}{K} in the configuration of $j_Q\parallel a$, $H\parallel b^*$ \cite{tu_Ba2Co(AsO4)2_2023}
            is also shown.
        }
        \label{comparison_honeycombs}
    \end{figure}

    Contrasting behavior between CTO and CNO highlights the diversity of MTC among honeycomb-lattice magnets. 
    Figure \ref{comparison_honeycombs} compares MTC ratios in honeycomb-lattice magnets at the temperature where the MTC ratio becomes the largest in the measured field range below \SI{15}{T}.
    Co-based materials show relatively large MTC change. 
    The MTC of $\mathrm{Na_2Co_2TeO_6}$ in the high magnetic fields is larger than that of the other honeycomb antiferromagnets.
    The magnetic energy gap extracted by a fitting of the thermal conductivity in $\mathrm{Na_2Co_2TeO_6}$ linearly increases in the high-field region ($\hbar\omega_0 \simeq \SI{5.4}{meV}$ at \SI{15}{T}), 
    thus suppressing phonon scattering due to magnetic excitation \cite{hong_Na2Co2TeO6_2021}.
    MTC in CTO is comparable to those in the other Co-based compounds, while CNO shows relatively small MTC. 
    Although a contrasting temperature-dependence of $\kappa_{xx}$ at zero field has been reported among trirutile compounds \cite{Prasai_AB2O6_2018}, 
    to the best of our knowledge, 
    it has been rarely reported and explored that such a wide variation of MTC happens within isostructural compounds with isoelectronic magnetic ions and similar magnetic and magnetoelastic properties.
    
    Although the analysis using Eq. (\ref{formula_omega}) succeeds in capturing the phenomenological understanding of the underlying physics observed, we note that it gives only
    becomes an indication of the presence of an energy scale of a gap of magnetic excitations relevant to phonon scattering. 
    The identification of the microscopic scattering process is a subject of future studies. 
    We also note that the magnetoelectric effect in CTO is known to be a nonmonotonic function of magnetic field, while CNO shows typical linear magnetoelectric behavior \cite{lee_highly_2020,khanh_manipulation_2017}. 
    This difference would also be linked to the distinct MTC through the field-induced evolution of magnetic structure and spin-lattice coupling. 
    In order to reveal these features and to deepen the understanding of the contrasting behaviors in CTO and CNO, further elastic/inelastic neutron diffraction studies are needed to investigate the spin structures in an in-plane magnetic field and to resolve the the magnetic excitation spectra particularly in CTO.

    In conclusion, a large enhancement of thermal conductivity in a magnetic field is observed in CTO in contrast to a much smaller magnetic-field-effect in CNO despite the apparent similarity between these compounds. 
    This discrepancy is attributed to the difference in the energy of magnetic excitations that scatter phonons and their field-evolution, which is consistent with relatively softer magnetization in CTO than that in CNO.
    These findings suggest a potential for the giant tunability of thermal conductivity in honeycomb-lattice magnets with a faint perturbation including the substitution of nonmagnetic elements. 
    This may also be relevant to harnessing exotic heat transport phenomena in Kitaev magnets.

    \section*{Acknowledgments}

    T.K. was supported by Ministry of Education Culture Sports Science and Technology (MEXT) Leading Initiative for Excellent Young Researchers (JPMXS0320200135).
    This study was supported by JSPS KAKENHI Grants-in-Aid (No. 19H01835, JP19H05826, 21K13874, 22K14010, and 23K13068), and Inamori Foundation.
    The synchrotron radiation experiments were performed at SPring-8 with the approval of the Japan Synchrotron Radiation Institute (JASRI) (Proposal No. 2022B1582). 
    This work was partly performed using the facilities of the Materials Design and Characterization Laboratory in the Institute for Solid State Physics, the University of Tokyo.
    The authors thank A. Ikeda for generously allowing the use of optical sensing instrument (Hyperion si155, LUNA) for thermal expansion/magnetostriction measurements.

    \nocite{Cao_Co4Nb2O9_2015,khanh_Co4Nb2O9_2016,Cao_Mn4Ta2O9_2018,lee_highly_2020}
    \nocite{Cao_Co4Nb2O9_2015,khanh_Co4Nb2O9_2016,Cao_Mn4Ta2O9_2018,lee_highly_2020}
    \nocite{Bertaut_A2B4O9_1961}
    \nocite{Castellanos_Co4Nb2O9_2006}
    \nocite{ding_Co4Nb2O9_2020,choi_noncollinear_2020}
    \nocite{khanh_Co4Nb2O9_2016,ding_Co4Nb2O9_2020,choi_noncollinear_2020}
    \nocite{ding_Co4Nb2O9_2020,choi_noncollinear_2020}
    \nocite{ding_Co4Nb2O9_2020,choi_noncollinear_2020}
    \nocite{ding_Co4Nb2O9_2020,choi_noncollinear_2020}
    \nocite{solovyev_origin_2016}
    \nocite{Xie_Co4Nb2O9_2016,Yin_Co4Nb2O9_2016,Khanh_Co4Nb2O9_2019}
    \nocite{Chang_Co4Nb2O9_2023}
    \nocite{khanh_Co4Nb2O9_2016,Yin_Co4Nb2O9_2016,lee_highly_2020}
    \nocite{Yin_Co4Nb2O9_2016,lee_highly_2020}
    \nocite{Khanh_Co4Nb2O9_2019,lee_highly_2020}
    \nocite{Chang_Co4Nb2O9_2023}
    \nocite{ding_Co4Nb2O9_2020}
    \nocite{khanh_manipulation_2017}
    \nocite{Cao_Co4Nb2O9_2015,khanh_Co4Nb2O9_2016,lee_highly_2020}
    \nocite{lee_highly_2020,choi_noncollinear_2020}
    \nocite{Callaway_CallawayModel_1959,Callaway_Callaway_1961}
    \nocite{Xie_Co4Nb2O9_2018}
    \nocite{Xie_Co4Ta2O9_2022}
    \nocite{SHEARD_phonon_1973,Wybourne_phonon_1985,Hentrich_RuCl3_2018,Prasai_AB2O6_2018,Hentrich_RuCl3_2020,yang_Cr2Si2Te6_2023}
    \nocite{Holland_thermalconductivity_1963}
    \nocite{Carruthers_thermalconductivity_1961}
    \nocite{narayanan_Mn4Ta2O9_2018}

    \bibliographystyle{apsrev4-2}

\begin{thebibliography}{91}%
\makeatletter
\providecommand \@ifxundefined [1]{%
 \@ifx{#1\undefined}
}%
\providecommand \@ifnum [1]{%
 \ifnum #1\expandafter \@firstoftwo
 \else \expandafter \@secondoftwo
 \fi
}%
\providecommand \@ifx [1]{%
 \ifx #1\expandafter \@firstoftwo
 \else \expandafter \@secondoftwo
 \fi
}%
\providecommand \natexlab [1]{#1}%
\providecommand \enquote  [1]{``#1''}%
\providecommand \bibnamefont  [1]{#1}%
\providecommand \bibfnamefont [1]{#1}%
\providecommand \citenamefont [1]{#1}%
\providecommand \href@noop [0]{\@secondoftwo}%
\providecommand \href [0]{\begingroup \@sanitize@url \@href}%
\providecommand \@href[1]{\@@startlink{#1}\@@href}%
\providecommand \@@href[1]{\endgroup#1\@@endlink}%
\providecommand \@sanitize@url [0]{\catcode `\\12\catcode `\$12\catcode `\&12\catcode `\#12\catcode `\^12\catcode `\_12\catcode `\%12\relax}%
\providecommand \@@startlink[1]{}%
\providecommand \@@endlink[0]{}%
\providecommand \url  [0]{\begingroup\@sanitize@url \@url }%
\providecommand \@url [1]{\endgroup\@href {#1}{\urlprefix }}%
\providecommand \urlprefix  [0]{URL }%
\providecommand \Eprint [0]{\href }%
\providecommand \doibase [0]{https://doi.org/}%
\providecommand \selectlanguage [0]{\@gobble}%
\providecommand \bibinfo  [0]{\@secondoftwo}%
\providecommand \bibfield  [0]{\@secondoftwo}%
\providecommand \translation [1]{[#1]}%
\providecommand \BibitemOpen [0]{}%
\providecommand \bibitemStop [0]{}%
\providecommand \bibitemNoStop [0]{.\EOS\space}%
\providecommand \EOS [0]{\spacefactor3000\relax}%
\providecommand \BibitemShut  [1]{\csname bibitem#1\endcsname}%
\let\auto@bib@innerbib\@empty
\bibitem [{\citenamefont {Li}\ and\ \citenamefont {Chen}(2020)}]{Li_probe_2020}%
  \BibitemOpen
  \bibfield  {author} {\bibinfo {author} {\bibfnamefont {M.}~\bibnamefont {Li}}\ and\ \bibinfo {author} {\bibfnamefont {G.}~\bibnamefont {Chen}},\ }\href {https://doi.org/10.1557/mrs.2020.124} {\bibfield  {journal} {\bibinfo  {journal} {MRS Bulletin}\ }\textbf {\bibinfo {volume} {45}},\ \bibinfo {pages} {348} (\bibinfo {year} {2020})}\BibitemShut {NoStop}%
\bibitem [{\citenamefont {Sologubenko}\ \emph {et~al.}(2007)\citenamefont {Sologubenko}, \citenamefont {Lorenz}, \citenamefont {Ott},\ and\ \citenamefont {Freimuth}}]{Sologubenko_excitation_2007}%
  \BibitemOpen
  \bibfield  {author} {\bibinfo {author} {\bibfnamefont {A.~V.}\ \bibnamefont {Sologubenko}}, \bibinfo {author} {\bibfnamefont {T.}~\bibnamefont {Lorenz}}, \bibinfo {author} {\bibfnamefont {H.~R.}\ \bibnamefont {Ott}},\ and\ \bibinfo {author} {\bibfnamefont {A.}~\bibnamefont {Freimuth}},\ }\href {https://doi.org/10.1007/s10909-007-9317-x} {\bibfield  {journal} {\bibinfo  {journal} {J. Low Temp. Phys.}\ }\textbf {\bibinfo {volume} {147}},\ \bibinfo {pages} {387} (\bibinfo {year} {2007})}\BibitemShut {NoStop}%
\bibitem [{\citenamefont {Hess}(2007)}]{Hess_heat_2007}%
  \BibitemOpen
  \bibfield  {author} {\bibinfo {author} {\bibfnamefont {C.}~\bibnamefont {Hess}},\ }\href {https://doi.org/10.1140/epjst/e2007-00363-8} {\bibfield  {journal} {\bibinfo  {journal} {Eur. Phys. J. Spec. Top.}\ }\textbf {\bibinfo {volume} {151}},\ \bibinfo {pages} {73} (\bibinfo {year} {2007})}\BibitemShut {NoStop}%
\bibitem [{\citenamefont {Strohm}\ \emph {et~al.}(2005)\citenamefont {Strohm}, \citenamefont {Rikken},\ and\ \citenamefont {Wyder}}]{Strohm_Tb3Ga5O12_2005}%
  \BibitemOpen
  \bibfield  {author} {\bibinfo {author} {\bibfnamefont {C.}~\bibnamefont {Strohm}}, \bibinfo {author} {\bibfnamefont {G.~L. J.~A.}\ \bibnamefont {Rikken}},\ and\ \bibinfo {author} {\bibfnamefont {P.}~\bibnamefont {Wyder}},\ }\href {https://doi.org/10.1103/PhysRevLett.95.155901} {\bibfield  {journal} {\bibinfo  {journal} {Phys. Rev. Lett.}\ }\textbf {\bibinfo {volume} {95}},\ \bibinfo {pages} {155901} (\bibinfo {year} {2005})}\BibitemShut {NoStop}%
\bibitem [{\citenamefont {Onose}\ \emph {et~al.}(2010)\citenamefont {Onose}, \citenamefont {Ideue}, \citenamefont {Katsura}, \citenamefont {Shiomi}, \citenamefont {Nagaosa},\ and\ \citenamefont {Tokura}}]{onose_observation_2010}%
  \BibitemOpen
  \bibfield  {author} {\bibinfo {author} {\bibfnamefont {Y.}~\bibnamefont {Onose}}, \bibinfo {author} {\bibfnamefont {T.}~\bibnamefont {Ideue}}, \bibinfo {author} {\bibfnamefont {H.}~\bibnamefont {Katsura}}, \bibinfo {author} {\bibfnamefont {Y.}~\bibnamefont {Shiomi}}, \bibinfo {author} {\bibfnamefont {N.}~\bibnamefont {Nagaosa}},\ and\ \bibinfo {author} {\bibfnamefont {Y.}~\bibnamefont {Tokura}},\ }\href {https://doi.org/10.1126/science.1188260} {\bibfield  {journal} {\bibinfo  {journal} {Science}\ }\textbf {\bibinfo {volume} {329}},\ \bibinfo {pages} {297} (\bibinfo {year} {2010})}\BibitemShut {NoStop}%
\bibitem [{\citenamefont {Grissonnanche}\ \emph {et~al.}(2019)\citenamefont {Grissonnanche}, \citenamefont {Legros}, \citenamefont {Badoux}, \citenamefont {Lefrançois}, \citenamefont {Zatko}, \citenamefont {Lizaire}, \citenamefont {Lalibert\'{e}}, \citenamefont {Gourgout}, \citenamefont {Zhou}, \citenamefont {Pyon}, \citenamefont {Takayama}, \citenamefont {Takagi}, \citenamefont {Ono}, \citenamefont {Doiron-Leyraud},\ and\ \citenamefont {Taillefer}}]{grissonnanche_giant_2019}%
  \BibitemOpen
  \bibfield  {author} {\bibinfo {author} {\bibfnamefont {G.}~\bibnamefont {Grissonnanche}}, \bibinfo {author} {\bibfnamefont {A.}~\bibnamefont {Legros}}, \bibinfo {author} {\bibfnamefont {S.}~\bibnamefont {Badoux}}, \bibinfo {author} {\bibfnamefont {E.}~\bibnamefont {Lefrançois}}, \bibinfo {author} {\bibfnamefont {V.}~\bibnamefont {Zatko}}, \bibinfo {author} {\bibfnamefont {M.}~\bibnamefont {Lizaire}}, \bibinfo {author} {\bibfnamefont {F.}~\bibnamefont {Lalibert\'{e}}}, \bibinfo {author} {\bibfnamefont {A.}~\bibnamefont {Gourgout}}, \bibinfo {author} {\bibfnamefont {J.-S.}\ \bibnamefont {Zhou}}, \bibinfo {author} {\bibfnamefont {S.}~\bibnamefont {Pyon}}, \bibinfo {author} {\bibfnamefont {T.}~\bibnamefont {Takayama}}, \bibinfo {author} {\bibfnamefont {H.}~\bibnamefont {Takagi}}, \bibinfo {author} {\bibfnamefont {S.}~\bibnamefont {Ono}}, \bibinfo {author} {\bibfnamefont {N.}~\bibnamefont {Doiron-Leyraud}},\ and\ \bibinfo {author} {\bibfnamefont {L.}~\bibnamefont {Taillefer}},\ }\href
  {https://doi.org/10.1038/s41586-019-1375-0} {\bibfield  {journal} {\bibinfo  {journal} {Nature}\ }\textbf {\bibinfo {volume} {571}},\ \bibinfo {pages} {376} (\bibinfo {year} {2019})}\BibitemShut {NoStop}%
\bibitem [{\citenamefont {Ando}\ \emph {et~al.}(1998)\citenamefont {Ando}, \citenamefont {Takeya}, \citenamefont {Sisson}, \citenamefont {Doettinger}, \citenamefont {Tanaka}, \citenamefont {Feigelson},\ and\ \citenamefont {Kapitulnik}}]{Ando_CuGeO3_1998}%
  \BibitemOpen
  \bibfield  {author} {\bibinfo {author} {\bibfnamefont {Y.}~\bibnamefont {Ando}}, \bibinfo {author} {\bibfnamefont {J.}~\bibnamefont {Takeya}}, \bibinfo {author} {\bibfnamefont {D.~L.}\ \bibnamefont {Sisson}}, \bibinfo {author} {\bibfnamefont {S.~G.}\ \bibnamefont {Doettinger}}, \bibinfo {author} {\bibfnamefont {I.}~\bibnamefont {Tanaka}}, \bibinfo {author} {\bibfnamefont {R.~S.}\ \bibnamefont {Feigelson}},\ and\ \bibinfo {author} {\bibfnamefont {A.}~\bibnamefont {Kapitulnik}},\ }\href {https://doi.org/10.1103/PhysRevB.58.R2913} {\bibfield  {journal} {\bibinfo  {journal} {Phys. Rev. B}\ }\textbf {\bibinfo {volume} {58}},\ \bibinfo {pages} {R2913} (\bibinfo {year} {1998})}\BibitemShut {NoStop}%
\bibitem [{\citenamefont {Hofmann}\ \emph {et~al.}(2001)\citenamefont {Hofmann}, \citenamefont {Lorenz}, \citenamefont {Uhrig}, \citenamefont {Kierspel}, \citenamefont {Zabara}, \citenamefont {Freimuth}, \citenamefont {Kageyama},\ and\ \citenamefont {Ueda}}]{Hofmann_SrCu2(BO3)2_2001}%
  \BibitemOpen
  \bibfield  {author} {\bibinfo {author} {\bibfnamefont {M.}~\bibnamefont {Hofmann}}, \bibinfo {author} {\bibfnamefont {T.}~\bibnamefont {Lorenz}}, \bibinfo {author} {\bibfnamefont {G.~S.}\ \bibnamefont {Uhrig}}, \bibinfo {author} {\bibfnamefont {H.}~\bibnamefont {Kierspel}}, \bibinfo {author} {\bibfnamefont {O.}~\bibnamefont {Zabara}}, \bibinfo {author} {\bibfnamefont {A.}~\bibnamefont {Freimuth}}, \bibinfo {author} {\bibfnamefont {H.}~\bibnamefont {Kageyama}},\ and\ \bibinfo {author} {\bibfnamefont {Y.}~\bibnamefont {Ueda}},\ }\href {https://doi.org/10.1103/PhysRevLett.87.047202} {\bibfield  {journal} {\bibinfo  {journal} {Phys. Rev. Lett.}\ }\textbf {\bibinfo {volume} {87}},\ \bibinfo {pages} {047202} (\bibinfo {year} {2001})}\BibitemShut {NoStop}%
\bibitem [{\citenamefont {Sales}\ \emph {et~al.}(2002)\citenamefont {Sales}, \citenamefont {Lumsden}, \citenamefont {Nagler}, \citenamefont {Mandrus},\ and\ \citenamefont {Jin}}]{sales_K2V3O8_2002}%
  \BibitemOpen
  \bibfield  {author} {\bibinfo {author} {\bibfnamefont {B.~C.}\ \bibnamefont {Sales}}, \bibinfo {author} {\bibfnamefont {M.~D.}\ \bibnamefont {Lumsden}}, \bibinfo {author} {\bibfnamefont {S.~E.}\ \bibnamefont {Nagler}}, \bibinfo {author} {\bibfnamefont {D.}~\bibnamefont {Mandrus}},\ and\ \bibinfo {author} {\bibfnamefont {R.}~\bibnamefont {Jin}},\ }\href {https://doi.org/10.1103/PhysRevLett.88.095901} {\bibfield  {journal} {\bibinfo  {journal} {Phys. Rev. Lett.}\ }\textbf {\bibinfo {volume} {88}},\ \bibinfo {pages} {095901} (\bibinfo {year} {2002})}\BibitemShut {NoStop}%
\bibitem [{\citenamefont {Chen}\ \emph {et~al.}(2011)\citenamefont {Chen}, \citenamefont {Wang}, \citenamefont {Ke}, \citenamefont {Zhao}, \citenamefont {Liu}, \citenamefont {Fan}, \citenamefont {Li}, \citenamefont {Zhao},\ and\ \citenamefont {Sun}}]{Chen_(CH3)2NH2CuCl3_2011}%
  \BibitemOpen
  \bibfield  {author} {\bibinfo {author} {\bibfnamefont {L.~M.}\ \bibnamefont {Chen}}, \bibinfo {author} {\bibfnamefont {X.~M.}\ \bibnamefont {Wang}}, \bibinfo {author} {\bibfnamefont {W.~P.}\ \bibnamefont {Ke}}, \bibinfo {author} {\bibfnamefont {Z.~Y.}\ \bibnamefont {Zhao}}, \bibinfo {author} {\bibfnamefont {X.~G.}\ \bibnamefont {Liu}}, \bibinfo {author} {\bibfnamefont {C.}~\bibnamefont {Fan}}, \bibinfo {author} {\bibfnamefont {Q.~J.}\ \bibnamefont {Li}}, \bibinfo {author} {\bibfnamefont {X.}~\bibnamefont {Zhao}},\ and\ \bibinfo {author} {\bibfnamefont {X.~F.}\ \bibnamefont {Sun}},\ }\href {https://doi.org/10.1103/PhysRevB.84.134429} {\bibfield  {journal} {\bibinfo  {journal} {Phys. Rev. B}\ }\textbf {\bibinfo {volume} {84}},\ \bibinfo {pages} {134429} (\bibinfo {year} {2011})}\BibitemShut {NoStop}%
\bibitem [{\citenamefont {Ke}\ \emph {et~al.}(2011)\citenamefont {Ke}, \citenamefont {Wang}, \citenamefont {Fan}, \citenamefont {Zhao}, \citenamefont {Liu}, \citenamefont {Chen}, \citenamefont {Li}, \citenamefont {Zhao},\ and\ \citenamefont {Sun}}]{Ke_Ba3Mn2O8_2011}%
  \BibitemOpen
  \bibfield  {author} {\bibinfo {author} {\bibfnamefont {W.~P.}\ \bibnamefont {Ke}}, \bibinfo {author} {\bibfnamefont {X.~M.}\ \bibnamefont {Wang}}, \bibinfo {author} {\bibfnamefont {C.}~\bibnamefont {Fan}}, \bibinfo {author} {\bibfnamefont {Z.~Y.}\ \bibnamefont {Zhao}}, \bibinfo {author} {\bibfnamefont {X.~G.}\ \bibnamefont {Liu}}, \bibinfo {author} {\bibfnamefont {L.~M.}\ \bibnamefont {Chen}}, \bibinfo {author} {\bibfnamefont {Q.~J.}\ \bibnamefont {Li}}, \bibinfo {author} {\bibfnamefont {X.}~\bibnamefont {Zhao}},\ and\ \bibinfo {author} {\bibfnamefont {X.~F.}\ \bibnamefont {Sun}},\ }\href {https://doi.org/10.1103/PhysRevB.84.094440} {\bibfield  {journal} {\bibinfo  {journal} {Phys. Rev. B}\ }\textbf {\bibinfo {volume} {84}},\ \bibinfo {pages} {094440} (\bibinfo {year} {2011})}\BibitemShut {NoStop}%
\bibitem [{\citenamefont {Wu}\ \emph {et~al.}(2016)\citenamefont {Wu}, \citenamefont {Song}, \citenamefont {Zhao}, \citenamefont {Shi}, \citenamefont {Xu}, \citenamefont {Zhao}, \citenamefont {Liu}, \citenamefont {Zhao},\ and\ \citenamefont {Sun}}]{wu_Cu3(CO3)2(OH)2_2016}%
  \BibitemOpen
  \bibfield  {author} {\bibinfo {author} {\bibfnamefont {J.~C.}\ \bibnamefont {Wu}}, \bibinfo {author} {\bibfnamefont {J.~D.}\ \bibnamefont {Song}}, \bibinfo {author} {\bibfnamefont {Z.~Y.}\ \bibnamefont {Zhao}}, \bibinfo {author} {\bibfnamefont {J.}~\bibnamefont {Shi}}, \bibinfo {author} {\bibfnamefont {H.~S.}\ \bibnamefont {Xu}}, \bibinfo {author} {\bibfnamefont {J.~Y.}\ \bibnamefont {Zhao}}, \bibinfo {author} {\bibfnamefont {X.~G.}\ \bibnamefont {Liu}}, \bibinfo {author} {\bibfnamefont {X.}~\bibnamefont {Zhao}},\ and\ \bibinfo {author} {\bibfnamefont {X.~F.}\ \bibnamefont {Sun}},\ }\href {https://doi.org/10.1088/0953-8984/28/5/056002} {\bibfield  {journal} {\bibinfo  {journal} {J. Phys.: Condens. Matter}\ }\textbf {\bibinfo {volume} {28}},\ \bibinfo {pages} {056002} (\bibinfo {year} {2016})}\BibitemShut {NoStop}%
\bibitem [{\citenamefont {Zhao}\ \emph {et~al.}(2016)\citenamefont {Zhao}, \citenamefont {Wu}, \citenamefont {Zhao}, \citenamefont {He}, \citenamefont {Song}, \citenamefont {Zhao}, \citenamefont {Liu}, \citenamefont {Sun},\ and\ \citenamefont {Li}}]{zhao_Co3V2O8_2016}%
  \BibitemOpen
  \bibfield  {author} {\bibinfo {author} {\bibfnamefont {X.}~\bibnamefont {Zhao}}, \bibinfo {author} {\bibfnamefont {J.~C.}\ \bibnamefont {Wu}}, \bibinfo {author} {\bibfnamefont {Z.~Y.}\ \bibnamefont {Zhao}}, \bibinfo {author} {\bibfnamefont {Z.~Z.}\ \bibnamefont {He}}, \bibinfo {author} {\bibfnamefont {J.~D.}\ \bibnamefont {Song}}, \bibinfo {author} {\bibfnamefont {J.~Y.}\ \bibnamefont {Zhao}}, \bibinfo {author} {\bibfnamefont {X.~G.}\ \bibnamefont {Liu}}, \bibinfo {author} {\bibfnamefont {X.~F.}\ \bibnamefont {Sun}},\ and\ \bibinfo {author} {\bibfnamefont {X.~G.}\ \bibnamefont {Li}},\ }\href {https://doi.org/10.1063/1.4953790} {\bibfield  {journal} {\bibinfo  {journal} {Appl. Phys. Lett.}\ }\textbf {\bibinfo {volume} {108}},\ \bibinfo {pages} {242405} (\bibinfo {year} {2016})}\BibitemShut {NoStop}%
\bibitem [{\citenamefont {Owerre}(2016{\natexlab{a}})}]{Owerre_theoretical_2016}%
  \BibitemOpen
  \bibfield  {author} {\bibinfo {author} {\bibfnamefont {S.~A.}\ \bibnamefont {Owerre}},\ }\href {https://doi.org/10.1088/0953-8984/28/38/386001} {\bibfield  {journal} {\bibinfo  {journal} {J. Phys.: Condens. Matter}\ }\textbf {\bibinfo {volume} {28}},\ \bibinfo {pages} {386001} (\bibinfo {year} {2016}{\natexlab{a}})}\BibitemShut {NoStop}%
\bibitem [{\citenamefont {Owerre}(2016{\natexlab{b}})}]{Owerre_honeycomb_2016}%
  \BibitemOpen
  \bibfield  {author} {\bibinfo {author} {\bibfnamefont {S.~A.}\ \bibnamefont {Owerre}},\ }\href {https://doi.org/10.1063/1.4959815} {\bibfield  {journal} {\bibinfo  {journal} {J. Appl. Phys.}\ }\textbf {\bibinfo {volume} {120}},\ \bibinfo {pages} {043903} (\bibinfo {year} {2016}{\natexlab{b}})}\BibitemShut {NoStop}%
\bibitem [{\citenamefont {Fransson}\ \emph {et~al.}(2016)\citenamefont {Fransson}, \citenamefont {Black-Schaffer},\ and\ \citenamefont {Balatsky}}]{Fransson_dirac_2016}%
  \BibitemOpen
  \bibfield  {author} {\bibinfo {author} {\bibfnamefont {J.}~\bibnamefont {Fransson}}, \bibinfo {author} {\bibfnamefont {A.~M.}\ \bibnamefont {Black-Schaffer}},\ and\ \bibinfo {author} {\bibfnamefont {A.~V.}\ \bibnamefont {Balatsky}},\ }\href {https://doi.org/10.1103/PhysRevB.94.075401} {\bibfield  {journal} {\bibinfo  {journal} {Phys. Rev. B}\ }\textbf {\bibinfo {volume} {94}},\ \bibinfo {pages} {075401} (\bibinfo {year} {2016})}\BibitemShut {NoStop}%
\bibitem [{\citenamefont {Kim}\ \emph {et~al.}(2016)\citenamefont {Kim}, \citenamefont {Ochoa}, \citenamefont {Zarzuela},\ and\ \citenamefont {Tserkovnyak}}]{Kim_HKMmodel_2016}%
  \BibitemOpen
  \bibfield  {author} {\bibinfo {author} {\bibfnamefont {S.~K.}\ \bibnamefont {Kim}}, \bibinfo {author} {\bibfnamefont {H.}~\bibnamefont {Ochoa}}, \bibinfo {author} {\bibfnamefont {R.}~\bibnamefont {Zarzuela}},\ and\ \bibinfo {author} {\bibfnamefont {Y.}~\bibnamefont {Tserkovnyak}},\ }\href {https://doi.org/10.1103/PhysRevLett.117.227201} {\bibfield  {journal} {\bibinfo  {journal} {Phys. Rev. Lett.}\ }\textbf {\bibinfo {volume} {117}},\ \bibinfo {pages} {227201} (\bibinfo {year} {2016})}\BibitemShut {NoStop}%
\bibitem [{\citenamefont {Chen}\ \emph {et~al.}(2018)\citenamefont {Chen}, \citenamefont {Chung}, \citenamefont {Gao}, \citenamefont {Chen}, \citenamefont {Stone}, \citenamefont {Kolesnikov}, \citenamefont {Huang},\ and\ \citenamefont {Dai}}]{Chen_CrI3_2018}%
  \BibitemOpen
  \bibfield  {author} {\bibinfo {author} {\bibfnamefont {L.}~\bibnamefont {Chen}}, \bibinfo {author} {\bibfnamefont {J.-H.}\ \bibnamefont {Chung}}, \bibinfo {author} {\bibfnamefont {B.}~\bibnamefont {Gao}}, \bibinfo {author} {\bibfnamefont {T.}~\bibnamefont {Chen}}, \bibinfo {author} {\bibfnamefont {M.~B.}\ \bibnamefont {Stone}}, \bibinfo {author} {\bibfnamefont {A.~I.}\ \bibnamefont {Kolesnikov}}, \bibinfo {author} {\bibfnamefont {Q.}~\bibnamefont {Huang}},\ and\ \bibinfo {author} {\bibfnamefont {P.}~\bibnamefont {Dai}},\ }\href {https://doi.org/10.1103/PhysRevX.8.041028} {\bibfield  {journal} {\bibinfo  {journal} {Phys. Rev. X}\ }\textbf {\bibinfo {volume} {8}},\ \bibinfo {pages} {041028} (\bibinfo {year} {2018})}\BibitemShut {NoStop}%
\bibitem [{\citenamefont {Yuan}\ \emph {et~al.}(2020)\citenamefont {Yuan}, \citenamefont {Khait}, \citenamefont {Shu}, \citenamefont {Chou}, \citenamefont {Stone}, \citenamefont {Clancy}, \citenamefont {Paramekanti},\ and\ \citenamefont {Kim}}]{Yuan_CoTiO3_2020}%
  \BibitemOpen
  \bibfield  {author} {\bibinfo {author} {\bibfnamefont {B.}~\bibnamefont {Yuan}}, \bibinfo {author} {\bibfnamefont {I.}~\bibnamefont {Khait}}, \bibinfo {author} {\bibfnamefont {G.-J.}\ \bibnamefont {Shu}}, \bibinfo {author} {\bibfnamefont {F.~C.}\ \bibnamefont {Chou}}, \bibinfo {author} {\bibfnamefont {M.~B.}\ \bibnamefont {Stone}}, \bibinfo {author} {\bibfnamefont {J.~P.}\ \bibnamefont {Clancy}}, \bibinfo {author} {\bibfnamefont {A.}~\bibnamefont {Paramekanti}},\ and\ \bibinfo {author} {\bibfnamefont {Y.-J.}\ \bibnamefont {Kim}},\ }\href {https://doi.org/10.1103/PhysRevX.10.011062} {\bibfield  {journal} {\bibinfo  {journal} {Phys. Rev. X}\ }\textbf {\bibinfo {volume} {10}},\ \bibinfo {pages} {011062} (\bibinfo {year} {2020})}\BibitemShut {NoStop}%
\bibitem [{\citenamefont {Zhu}\ \emph {et~al.}(2021)\citenamefont {Zhu}, \citenamefont {Zhang}, \citenamefont {Wang}, \citenamefont {dos Santos}, \citenamefont {Song}, \citenamefont {Mueller}, \citenamefont {Schmalzl}, \citenamefont {Schmidt}, \citenamefont {Ivanov}, \citenamefont {Park}, \citenamefont {Xu}, \citenamefont {Ma}, \citenamefont {Lounis}, \citenamefont {Blügel}, \citenamefont {Mokrousov}, \citenamefont {Su},\ and\ \citenamefont {Brückel}}]{Fengfeng_CrSiTe3CrGeTe3_2021}%
  \BibitemOpen
  \bibfield  {author} {\bibinfo {author} {\bibfnamefont {F.}~\bibnamefont {Zhu}}, \bibinfo {author} {\bibfnamefont {L.}~\bibnamefont {Zhang}}, \bibinfo {author} {\bibfnamefont {X.}~\bibnamefont {Wang}}, \bibinfo {author} {\bibfnamefont {F.~J.}\ \bibnamefont {dos Santos}}, \bibinfo {author} {\bibfnamefont {J.}~\bibnamefont {Song}}, \bibinfo {author} {\bibfnamefont {T.}~\bibnamefont {Mueller}}, \bibinfo {author} {\bibfnamefont {K.}~\bibnamefont {Schmalzl}}, \bibinfo {author} {\bibfnamefont {W.~F.}\ \bibnamefont {Schmidt}}, \bibinfo {author} {\bibfnamefont {A.}~\bibnamefont {Ivanov}}, \bibinfo {author} {\bibfnamefont {J.~T.}\ \bibnamefont {Park}}, \bibinfo {author} {\bibfnamefont {J.}~\bibnamefont {Xu}}, \bibinfo {author} {\bibfnamefont {J.}~\bibnamefont {Ma}}, \bibinfo {author} {\bibfnamefont {S.}~\bibnamefont {Lounis}}, \bibinfo {author} {\bibfnamefont {S.}~\bibnamefont {Blügel}}, \bibinfo {author} {\bibfnamefont {Y.}~\bibnamefont {Mokrousov}}, \bibinfo {author} {\bibfnamefont {Y.}~\bibnamefont {Su}},\ and\
  \bibinfo {author} {\bibfnamefont {T.}~\bibnamefont {Brückel}},\ }\href {https://doi.org/10.1126/sciadv.abi7532} {\bibfield  {journal} {\bibinfo  {journal} {Sci. Adv.}\ }\textbf {\bibinfo {volume} {7}},\ \bibinfo {pages} {eabi7532} (\bibinfo {year} {2021})}\BibitemShut {NoStop}%
\bibitem [{\citenamefont {Kitaev}(2006)}]{kitaev_anyons_2006}%
  \BibitemOpen
  \bibfield  {author} {\bibinfo {author} {\bibfnamefont {A.}~\bibnamefont {Kitaev}},\ }\href {https://doi.org/10.1016/j.aop.2005.10.005} {\bibfield  {journal} {\bibinfo  {journal} {Ann. Phys.}\ }\textbf {\bibinfo {volume} {321}},\ \bibinfo {pages} {2} (\bibinfo {year} {2006})}\BibitemShut {NoStop}%
\bibitem [{\citenamefont {Kasahara}\ \emph {et~al.}(2018{\natexlab{a}})\citenamefont {Kasahara}, \citenamefont {Ohnishi}, \citenamefont {Mizukami}, \citenamefont {Tanaka}, \citenamefont {Ma}, \citenamefont {Sugii}, \citenamefont {Kurita}, \citenamefont {Tanaka}, \citenamefont {Nasu}, \citenamefont {Motome}, \citenamefont {Shibauchi},\ and\ \citenamefont {Matsuda}}]{kasahara_RuCl3_2018}%
  \BibitemOpen
  \bibfield  {author} {\bibinfo {author} {\bibfnamefont {Y.}~\bibnamefont {Kasahara}}, \bibinfo {author} {\bibfnamefont {T.}~\bibnamefont {Ohnishi}}, \bibinfo {author} {\bibfnamefont {Y.}~\bibnamefont {Mizukami}}, \bibinfo {author} {\bibfnamefont {O.}~\bibnamefont {Tanaka}}, \bibinfo {author} {\bibfnamefont {S.}~\bibnamefont {Ma}}, \bibinfo {author} {\bibfnamefont {K.}~\bibnamefont {Sugii}}, \bibinfo {author} {\bibfnamefont {N.}~\bibnamefont {Kurita}}, \bibinfo {author} {\bibfnamefont {H.}~\bibnamefont {Tanaka}}, \bibinfo {author} {\bibfnamefont {J.}~\bibnamefont {Nasu}}, \bibinfo {author} {\bibfnamefont {Y.}~\bibnamefont {Motome}}, \bibinfo {author} {\bibfnamefont {T.}~\bibnamefont {Shibauchi}},\ and\ \bibinfo {author} {\bibfnamefont {Y.}~\bibnamefont {Matsuda}},\ }\href {https://doi.org/10.1038/s41586-018-0274-0} {\bibfield  {journal} {\bibinfo  {journal} {Nature}\ }\textbf {\bibinfo {volume} {559}},\ \bibinfo {pages} {227} (\bibinfo {year} {2018}{\natexlab{a}})}\BibitemShut {NoStop}%
\bibitem [{\citenamefont {Zhang}\ and\ \citenamefont {Niu}(2015)}]{Zhang_chiralphonon_2015}%
  \BibitemOpen
  \bibfield  {author} {\bibinfo {author} {\bibfnamefont {L.}~\bibnamefont {Zhang}}\ and\ \bibinfo {author} {\bibfnamefont {Q.}~\bibnamefont {Niu}},\ }\href {https://doi.org/10.1103/PhysRevLett.115.115502} {\bibfield  {journal} {\bibinfo  {journal} {Phys. Rev. Lett.}\ }\textbf {\bibinfo {volume} {115}},\ \bibinfo {pages} {115502} (\bibinfo {year} {2015})}\BibitemShut {NoStop}%
\bibitem [{\citenamefont {Zhu}\ \emph {et~al.}(2018)\citenamefont {Zhu}, \citenamefont {Yi}, \citenamefont {Li}, \citenamefont {Xiao}, \citenamefont {Zhang}, \citenamefont {Yang}, \citenamefont {Kaindl}, \citenamefont {Li}, \citenamefont {Wang},\ and\ \citenamefont {Zhang}}]{Hanyu_chiralphonon_2018}%
  \BibitemOpen
  \bibfield  {author} {\bibinfo {author} {\bibfnamefont {H.}~\bibnamefont {Zhu}}, \bibinfo {author} {\bibfnamefont {J.}~\bibnamefont {Yi}}, \bibinfo {author} {\bibfnamefont {M.-Y.}\ \bibnamefont {Li}}, \bibinfo {author} {\bibfnamefont {J.}~\bibnamefont {Xiao}}, \bibinfo {author} {\bibfnamefont {L.}~\bibnamefont {Zhang}}, \bibinfo {author} {\bibfnamefont {C.-W.}\ \bibnamefont {Yang}}, \bibinfo {author} {\bibfnamefont {R.~A.}\ \bibnamefont {Kaindl}}, \bibinfo {author} {\bibfnamefont {L.-J.}\ \bibnamefont {Li}}, \bibinfo {author} {\bibfnamefont {Y.}~\bibnamefont {Wang}},\ and\ \bibinfo {author} {\bibfnamefont {X.}~\bibnamefont {Zhang}},\ }\href {https://doi.org/10.1126/science.aar2711} {\bibfield  {journal} {\bibinfo  {journal} {Science}\ }\textbf {\bibinfo {volume} {359}},\ \bibinfo {pages} {579} (\bibinfo {year} {2018})}\BibitemShut {NoStop}%
\bibitem [{\citenamefont {Wu}\ \emph {et~al.}(2023)\citenamefont {Wu}, \citenamefont {Bao}, \citenamefont {Zhou}, \citenamefont {Wang}, \citenamefont {Sun}, \citenamefont {Wen}, \citenamefont {Wan},\ and\ \citenamefont {Zhang}}]{Wu_chiralphonon_2023}%
  \BibitemOpen
  \bibfield  {author} {\bibinfo {author} {\bibfnamefont {F.}~\bibnamefont {Wu}}, \bibinfo {author} {\bibfnamefont {S.}~\bibnamefont {Bao}}, \bibinfo {author} {\bibfnamefont {J.}~\bibnamefont {Zhou}}, \bibinfo {author} {\bibfnamefont {Y.}~\bibnamefont {Wang}}, \bibinfo {author} {\bibfnamefont {J.}~\bibnamefont {Sun}}, \bibinfo {author} {\bibfnamefont {J.}~\bibnamefont {Wen}}, \bibinfo {author} {\bibfnamefont {Y.}~\bibnamefont {Wan}},\ and\ \bibinfo {author} {\bibfnamefont {Q.}~\bibnamefont {Zhang}},\ }\href {https://doi.org/10.1038/s41567-023-02210-4} {\bibfield  {journal} {\bibinfo  {journal} {Nat. Phys.}\ } (\bibinfo {year} {2023})},\ \Eprint {https://arxiv.org/abs/https://doi.org/10.1038/s41567-023-02210-4} {https://doi.org/10.1038/s41567-023-02210-4} \BibitemShut {NoStop}%
\bibitem [{\citenamefont {Leahy}\ \emph {et~al.}(2017)\citenamefont {Leahy}, \citenamefont {Pocs}, \citenamefont {Siegfried}, \citenamefont {Graf}, \citenamefont {Do}, \citenamefont {Choi}, \citenamefont {Normand},\ and\ \citenamefont {Lee}}]{Leahy_RuCl3_2017}%
  \BibitemOpen
  \bibfield  {author} {\bibinfo {author} {\bibfnamefont {I.~A.}\ \bibnamefont {Leahy}}, \bibinfo {author} {\bibfnamefont {C.~A.}\ \bibnamefont {Pocs}}, \bibinfo {author} {\bibfnamefont {P.~E.}\ \bibnamefont {Siegfried}}, \bibinfo {author} {\bibfnamefont {D.}~\bibnamefont {Graf}}, \bibinfo {author} {\bibfnamefont {S.-H.}\ \bibnamefont {Do}}, \bibinfo {author} {\bibfnamefont {K.-Y.}\ \bibnamefont {Choi}}, \bibinfo {author} {\bibfnamefont {B.}~\bibnamefont {Normand}},\ and\ \bibinfo {author} {\bibfnamefont {M.}~\bibnamefont {Lee}},\ }\href {https://doi.org/10.1103/PhysRevLett.118.187203} {\bibfield  {journal} {\bibinfo  {journal} {Phys. Rev. Lett.}\ }\textbf {\bibinfo {volume} {118}},\ \bibinfo {pages} {187203} (\bibinfo {year} {2017})}\BibitemShut {NoStop}%
\bibitem [{\citenamefont {Pocs}\ \emph {et~al.}(2020)\citenamefont {Pocs}, \citenamefont {Leahy}, \citenamefont {Zheng}, \citenamefont {Cao}, \citenamefont {Choi}, \citenamefont {Do}, \citenamefont {Choi}, \citenamefont {Normand},\ and\ \citenamefont {Lee}}]{Pocs_CrCl3_2020}%
  \BibitemOpen
  \bibfield  {author} {\bibinfo {author} {\bibfnamefont {C.~A.}\ \bibnamefont {Pocs}}, \bibinfo {author} {\bibfnamefont {I.~A.}\ \bibnamefont {Leahy}}, \bibinfo {author} {\bibfnamefont {H.}~\bibnamefont {Zheng}}, \bibinfo {author} {\bibfnamefont {G.}~\bibnamefont {Cao}}, \bibinfo {author} {\bibfnamefont {E.-S.}\ \bibnamefont {Choi}}, \bibinfo {author} {\bibfnamefont {S.-H.}\ \bibnamefont {Do}}, \bibinfo {author} {\bibfnamefont {K.-Y.}\ \bibnamefont {Choi}}, \bibinfo {author} {\bibfnamefont {B.}~\bibnamefont {Normand}},\ and\ \bibinfo {author} {\bibfnamefont {M.}~\bibnamefont {Lee}},\ }\href {https://doi.org/10.1103/PhysRevResearch.2.013059} {\bibfield  {journal} {\bibinfo  {journal} {Phys. Rev. Res.}\ }\textbf {\bibinfo {volume} {2}},\ \bibinfo {pages} {013059} (\bibinfo {year} {2020})}\BibitemShut {NoStop}%
\bibitem [{\citenamefont {Zhong}\ \emph {et~al.}(2020)\citenamefont {Zhong}, \citenamefont {Gao}, \citenamefont {Ong},\ and\ \citenamefont {Cava}}]{zhong_Ba2Co(AsO4)2_2020}%
  \BibitemOpen
  \bibfield  {author} {\bibinfo {author} {\bibfnamefont {R.}~\bibnamefont {Zhong}}, \bibinfo {author} {\bibfnamefont {T.}~\bibnamefont {Gao}}, \bibinfo {author} {\bibfnamefont {N.~P.}\ \bibnamefont {Ong}},\ and\ \bibinfo {author} {\bibfnamefont {R.~J.}\ \bibnamefont {Cava}},\ }\href {https://doi.org/10.1126/sciadv.aay6953} {\bibfield  {journal} {\bibinfo  {journal} {Sci. Adv.}\ }\textbf {\bibinfo {volume} {6}},\ \bibinfo {pages} {eaay6953} (\bibinfo {year} {2020})}\BibitemShut {NoStop}%
\bibitem [{\citenamefont {Hong}\ \emph {et~al.}(2021)\citenamefont {Hong}, \citenamefont {Gillig}, \citenamefont {Hentrich}, \citenamefont {Yao}, \citenamefont {Kocsis}, \citenamefont {Witte}, \citenamefont {Schreiner}, \citenamefont {Baumann}, \citenamefont {P\'erez}, \citenamefont {Wolter}, \citenamefont {Li}, \citenamefont {B\"uchner},\ and\ \citenamefont {Hess}}]{hong_Na2Co2TeO6_2021}%
  \BibitemOpen
  \bibfield  {author} {\bibinfo {author} {\bibfnamefont {X.}~\bibnamefont {Hong}}, \bibinfo {author} {\bibfnamefont {M.}~\bibnamefont {Gillig}}, \bibinfo {author} {\bibfnamefont {R.}~\bibnamefont {Hentrich}}, \bibinfo {author} {\bibfnamefont {W.}~\bibnamefont {Yao}}, \bibinfo {author} {\bibfnamefont {V.}~\bibnamefont {Kocsis}}, \bibinfo {author} {\bibfnamefont {A.~R.}\ \bibnamefont {Witte}}, \bibinfo {author} {\bibfnamefont {T.}~\bibnamefont {Schreiner}}, \bibinfo {author} {\bibfnamefont {D.}~\bibnamefont {Baumann}}, \bibinfo {author} {\bibfnamefont {N.}~\bibnamefont {P\'erez}}, \bibinfo {author} {\bibfnamefont {A.~U.~B.}\ \bibnamefont {Wolter}}, \bibinfo {author} {\bibfnamefont {Y.}~\bibnamefont {Li}}, \bibinfo {author} {\bibfnamefont {B.}~\bibnamefont {B\"uchner}},\ and\ \bibinfo {author} {\bibfnamefont {C.}~\bibnamefont {Hess}},\ }\href {https://doi.org/10.1103/PhysRevB.104.144426} {\bibfield  {journal} {\bibinfo  {journal} {Phys. Rev. B}\ }\textbf {\bibinfo {volume} {104}},\ \bibinfo {pages} {144426}
  (\bibinfo {year} {2021})}\BibitemShut {NoStop}%
\bibitem [{\citenamefont {Tu}\ \emph {et~al.}(2023)\citenamefont {Tu}, \citenamefont {Dai}, \citenamefont {Zhang}, \citenamefont {Zhao}, \citenamefont {Jin}, \citenamefont {Gao}, \citenamefont {Chen}, \citenamefont {Dai},\ and\ \citenamefont {Li}}]{tu_Ba2Co(AsO4)2_2023}%
  \BibitemOpen
  \bibfield  {author} {\bibinfo {author} {\bibfnamefont {C.}~\bibnamefont {Tu}}, \bibinfo {author} {\bibfnamefont {D.}~\bibnamefont {Dai}}, \bibinfo {author} {\bibfnamefont {X.}~\bibnamefont {Zhang}}, \bibinfo {author} {\bibfnamefont {C.}~\bibnamefont {Zhao}}, \bibinfo {author} {\bibfnamefont {X.}~\bibnamefont {Jin}}, \bibinfo {author} {\bibfnamefont {B.}~\bibnamefont {Gao}}, \bibinfo {author} {\bibfnamefont {T.}~\bibnamefont {Chen}}, \bibinfo {author} {\bibfnamefont {P.}~\bibnamefont {Dai}},\ and\ \bibinfo {author} {\bibfnamefont {S.}~\bibnamefont {Li}},\ }\href@noop {} {} (\bibinfo {year} {2023}),\ \Eprint {https://arxiv.org/abs/2212.07322} {arXiv:2212.07322 [cond-mat.str-el]} \BibitemShut {NoStop}%
\bibitem [{\citenamefont {Ideue}\ \emph {et~al.}(2017)\citenamefont {Ideue}, \citenamefont {Kurumaji}, \citenamefont {Ishiwata},\ and\ \citenamefont {Tokura}}]{ideue_Fe2Mo3O8_2017}%
  \BibitemOpen
  \bibfield  {author} {\bibinfo {author} {\bibfnamefont {T.}~\bibnamefont {Ideue}}, \bibinfo {author} {\bibfnamefont {T.}~\bibnamefont {Kurumaji}}, \bibinfo {author} {\bibfnamefont {S.}~\bibnamefont {Ishiwata}},\ and\ \bibinfo {author} {\bibfnamefont {Y.}~\bibnamefont {Tokura}},\ }\href {https://doi.org/10.1038/nmat4905} {\bibfield  {journal} {\bibinfo  {journal} {Nat. Mater.}\ }\textbf {\bibinfo {volume} {16}},\ \bibinfo {pages} {797} (\bibinfo {year} {2017})}\BibitemShut {NoStop}%
\bibitem [{\citenamefont {Kasahara}\ \emph {et~al.}(2018{\natexlab{b}})\citenamefont {Kasahara}, \citenamefont {Sugii}, \citenamefont {Ohnishi}, \citenamefont {Shimozawa}, \citenamefont {Yamashita}, \citenamefont {Kurita}, \citenamefont {Tanaka}, \citenamefont {Nasu}, \citenamefont {Motome}, \citenamefont {Shibauchi},\ and\ \citenamefont {Matsuda}}]{Kasahara_RuCl3Hall_2018}%
  \BibitemOpen
  \bibfield  {author} {\bibinfo {author} {\bibfnamefont {Y.}~\bibnamefont {Kasahara}}, \bibinfo {author} {\bibfnamefont {K.}~\bibnamefont {Sugii}}, \bibinfo {author} {\bibfnamefont {T.}~\bibnamefont {Ohnishi}}, \bibinfo {author} {\bibfnamefont {M.}~\bibnamefont {Shimozawa}}, \bibinfo {author} {\bibfnamefont {M.}~\bibnamefont {Yamashita}}, \bibinfo {author} {\bibfnamefont {N.}~\bibnamefont {Kurita}}, \bibinfo {author} {\bibfnamefont {H.}~\bibnamefont {Tanaka}}, \bibinfo {author} {\bibfnamefont {J.}~\bibnamefont {Nasu}}, \bibinfo {author} {\bibfnamefont {Y.}~\bibnamefont {Motome}}, \bibinfo {author} {\bibfnamefont {T.}~\bibnamefont {Shibauchi}},\ and\ \bibinfo {author} {\bibfnamefont {Y.}~\bibnamefont {Matsuda}},\ }\href {https://doi.org/10.1103/PhysRevLett.120.217205} {\bibfield  {journal} {\bibinfo  {journal} {Phys. Rev. Lett.}\ }\textbf {\bibinfo {volume} {120}},\ \bibinfo {pages} {217205} (\bibinfo {year} {2018}{\natexlab{b}})}\BibitemShut {NoStop}%
\bibitem [{\citenamefont {Zhang}\ \emph {et~al.}(2021)\citenamefont {Zhang}, \citenamefont {Xu}, \citenamefont {Carnahan}, \citenamefont {Sretenovic}, \citenamefont {Suri}, \citenamefont {Xiao},\ and\ \citenamefont {Ke}}]{zhang_VI3_2021}%
  \BibitemOpen
  \bibfield  {author} {\bibinfo {author} {\bibfnamefont {H.}~\bibnamefont {Zhang}}, \bibinfo {author} {\bibfnamefont {C.}~\bibnamefont {Xu}}, \bibinfo {author} {\bibfnamefont {C.}~\bibnamefont {Carnahan}}, \bibinfo {author} {\bibfnamefont {M.}~\bibnamefont {Sretenovic}}, \bibinfo {author} {\bibfnamefont {N.}~\bibnamefont {Suri}}, \bibinfo {author} {\bibfnamefont {D.}~\bibnamefont {Xiao}},\ and\ \bibinfo {author} {\bibfnamefont {X.}~\bibnamefont {Ke}},\ }\href {https://doi.org/10.1103/PhysRevLett.127.247202} {\bibfield  {journal} {\bibinfo  {journal} {Phys. Rev. Lett.}\ }\textbf {\bibinfo {volume} {127}},\ \bibinfo {pages} {247202} (\bibinfo {year} {2021})}\BibitemShut {NoStop}%
\bibitem [{\citenamefont {Yang}\ \emph {et~al.}(2022)\citenamefont {Yang}, \citenamefont {Xu}, \citenamefont {Lee}, \citenamefont {Oh}, \citenamefont {Cheong},\ and\ \citenamefont {Park}}]{yang_Ni3TeO6_2022}%
  \BibitemOpen
  \bibfield  {author} {\bibinfo {author} {\bibfnamefont {H.}~\bibnamefont {Yang}}, \bibinfo {author} {\bibfnamefont {X.}~\bibnamefont {Xu}}, \bibinfo {author} {\bibfnamefont {J.~H.}\ \bibnamefont {Lee}}, \bibinfo {author} {\bibfnamefont {Y.~S.}\ \bibnamefont {Oh}}, \bibinfo {author} {\bibfnamefont {S.-W.}\ \bibnamefont {Cheong}},\ and\ \bibinfo {author} {\bibfnamefont {J.-G.}\ \bibnamefont {Park}},\ }\href {https://doi.org/10.1103/PhysRevB.106.144417} {\bibfield  {journal} {\bibinfo  {journal} {Phys. Rev. B}\ }\textbf {\bibinfo {volume} {106}},\ \bibinfo {pages} {144417} (\bibinfo {year} {2022})}\BibitemShut {NoStop}%
\bibitem [{\citenamefont {Choi}\ \emph {et~al.}(2023)\citenamefont {Choi}, \citenamefont {Yang}, \citenamefont {Park},\ and\ \citenamefont {Park}}]{Choi_Cr2Ge2Te6_2023}%
  \BibitemOpen
  \bibfield  {author} {\bibinfo {author} {\bibfnamefont {Y.}~\bibnamefont {Choi}}, \bibinfo {author} {\bibfnamefont {H.}~\bibnamefont {Yang}}, \bibinfo {author} {\bibfnamefont {J.}~\bibnamefont {Park}},\ and\ \bibinfo {author} {\bibfnamefont {J.-G.}\ \bibnamefont {Park}},\ }\href {https://doi.org/10.1103/PhysRevB.107.184434} {\bibfield  {journal} {\bibinfo  {journal} {Phys. Rev. B}\ }\textbf {\bibinfo {volume} {107}},\ \bibinfo {pages} {184434} (\bibinfo {year} {2023})}\BibitemShut {NoStop}%
\bibitem [{\citenamefont {Chern}\ \emph {et~al.}(2021)\citenamefont {Chern}, \citenamefont {Zhang},\ and\ \citenamefont {Kim}}]{Chern_sign_2021}%
  \BibitemOpen
  \bibfield  {author} {\bibinfo {author} {\bibfnamefont {L.~E.}\ \bibnamefont {Chern}}, \bibinfo {author} {\bibfnamefont {E.~Z.}\ \bibnamefont {Zhang}},\ and\ \bibinfo {author} {\bibfnamefont {Y.~B.}\ \bibnamefont {Kim}},\ }\href {https://doi.org/10.1103/PhysRevLett.126.147201} {\bibfield  {journal} {\bibinfo  {journal} {Phys. Rev. Lett.}\ }\textbf {\bibinfo {volume} {126}},\ \bibinfo {pages} {147201} (\bibinfo {year} {2021})}\BibitemShut {NoStop}%
\bibitem [{\citenamefont {Utermohlen}\ and\ \citenamefont {Trivedi}(2021)}]{Utermohlen_tensor_2021}%
  \BibitemOpen
  \bibfield  {author} {\bibinfo {author} {\bibfnamefont {F.~G.}\ \bibnamefont {Utermohlen}}\ and\ \bibinfo {author} {\bibfnamefont {N.}~\bibnamefont {Trivedi}},\ }\href {https://doi.org/10.1103/PhysRevB.103.155124} {\bibfield  {journal} {\bibinfo  {journal} {Phys. Rev. B}\ }\textbf {\bibinfo {volume} {103}},\ \bibinfo {pages} {155124} (\bibinfo {year} {2021})}\BibitemShut {NoStop}%
\bibitem [{\citenamefont {Yokoi}\ \emph {et~al.}(2021)\citenamefont {Yokoi}, \citenamefont {Ma}, \citenamefont {Kasahara}, \citenamefont {Kasahara}, \citenamefont {Shibauchi}, \citenamefont {Kurita}, \citenamefont {Tanaka}, \citenamefont {Nasu}, \citenamefont {Motome}, \citenamefont {Hickey}, \citenamefont {Trebst},\ and\ \citenamefont {Matsuda}}]{Yokoi_RuCl3_2021}%
  \BibitemOpen
  \bibfield  {author} {\bibinfo {author} {\bibfnamefont {T.}~\bibnamefont {Yokoi}}, \bibinfo {author} {\bibfnamefont {S.}~\bibnamefont {Ma}}, \bibinfo {author} {\bibfnamefont {Y.}~\bibnamefont {Kasahara}}, \bibinfo {author} {\bibfnamefont {S.}~\bibnamefont {Kasahara}}, \bibinfo {author} {\bibfnamefont {T.}~\bibnamefont {Shibauchi}}, \bibinfo {author} {\bibfnamefont {N.}~\bibnamefont {Kurita}}, \bibinfo {author} {\bibfnamefont {H.}~\bibnamefont {Tanaka}}, \bibinfo {author} {\bibfnamefont {J.}~\bibnamefont {Nasu}}, \bibinfo {author} {\bibfnamefont {Y.}~\bibnamefont {Motome}}, \bibinfo {author} {\bibfnamefont {C.}~\bibnamefont {Hickey}}, \bibinfo {author} {\bibfnamefont {S.}~\bibnamefont {Trebst}},\ and\ \bibinfo {author} {\bibfnamefont {Y.}~\bibnamefont {Matsuda}},\ }\href {https://doi.org/10.1126/science.aay5551} {\bibfield  {journal} {\bibinfo  {journal} {Science}\ }\textbf {\bibinfo {volume} {373}},\ \bibinfo {pages} {568} (\bibinfo {year} {2021})}\BibitemShut {NoStop}%
\bibitem [{\citenamefont {Takeda}\ \emph {et~al.}(2022)\citenamefont {Takeda}, \citenamefont {Mai}, \citenamefont {Akazawa}, \citenamefont {Tamura}, \citenamefont {Yan}, \citenamefont {Moovendaran}, \citenamefont {Raju}, \citenamefont {Sankar}, \citenamefont {Choi},\ and\ \citenamefont {Yamashita}}]{takeda_Na2Co2TeO6_2022}%
  \BibitemOpen
  \bibfield  {author} {\bibinfo {author} {\bibfnamefont {H.}~\bibnamefont {Takeda}}, \bibinfo {author} {\bibfnamefont {J.}~\bibnamefont {Mai}}, \bibinfo {author} {\bibfnamefont {M.}~\bibnamefont {Akazawa}}, \bibinfo {author} {\bibfnamefont {K.}~\bibnamefont {Tamura}}, \bibinfo {author} {\bibfnamefont {J.}~\bibnamefont {Yan}}, \bibinfo {author} {\bibfnamefont {K.}~\bibnamefont {Moovendaran}}, \bibinfo {author} {\bibfnamefont {K.}~\bibnamefont {Raju}}, \bibinfo {author} {\bibfnamefont {R.}~\bibnamefont {Sankar}}, \bibinfo {author} {\bibfnamefont {K.-Y.}\ \bibnamefont {Choi}},\ and\ \bibinfo {author} {\bibfnamefont {M.}~\bibnamefont {Yamashita}},\ }\href {https://doi.org/10.1103/PhysRevResearch.4.L042035} {\bibfield  {journal} {\bibinfo  {journal} {Phys. Rev. Res.}\ }\textbf {\bibinfo {volume} {4}},\ \bibinfo {pages} {L042035} (\bibinfo {year} {2022})}\BibitemShut {NoStop}%
\bibitem [{\citenamefont {Czajka}\ \emph {et~al.}(2023)\citenamefont {Czajka}, \citenamefont {Gao}, \citenamefont {Hirschberger}, \citenamefont {Lampen-Kelley}, \citenamefont {Banerjee}, \citenamefont {Quirk}, \citenamefont {Mandrus}, \citenamefont {Nagler},\ and\ \citenamefont {Ong}}]{Czajka_RuCl3_2023}%
  \BibitemOpen
  \bibfield  {author} {\bibinfo {author} {\bibfnamefont {P.}~\bibnamefont {Czajka}}, \bibinfo {author} {\bibfnamefont {T.}~\bibnamefont {Gao}}, \bibinfo {author} {\bibfnamefont {M.}~\bibnamefont {Hirschberger}}, \bibinfo {author} {\bibfnamefont {P.}~\bibnamefont {Lampen-Kelley}}, \bibinfo {author} {\bibfnamefont {A.}~\bibnamefont {Banerjee}}, \bibinfo {author} {\bibfnamefont {N.}~\bibnamefont {Quirk}}, \bibinfo {author} {\bibfnamefont {D.~G.}\ \bibnamefont {Mandrus}}, \bibinfo {author} {\bibfnamefont {S.~E.}\ \bibnamefont {Nagler}},\ and\ \bibinfo {author} {\bibfnamefont {N.~P.}\ \bibnamefont {Ong}},\ }\href {https://doi.org/10.1038/s41563-022-01397-w} {\bibfield  {journal} {\bibinfo  {journal} {Nat. Mater.}\ }\textbf {\bibinfo {volume} {22}},\ \bibinfo {pages} {36} (\bibinfo {year} {2023})}\BibitemShut {NoStop}%
\bibitem [{\citenamefont {Kurumaji}(2023)}]{Kurumaji_symmetry_2023}%
  \BibitemOpen
  \bibfield  {author} {\bibinfo {author} {\bibfnamefont {T.}~\bibnamefont {Kurumaji}},\ }\href {https://doi.org/10.1103/PhysRevResearch.5.023138} {\bibfield  {journal} {\bibinfo  {journal} {Phys. Rev. Res.}\ }\textbf {\bibinfo {volume} {5}},\ \bibinfo {pages} {023138} (\bibinfo {year} {2023})}\BibitemShut {NoStop}%
\bibitem [{\citenamefont {Chen}\ \emph {et~al.}(2023)\citenamefont {Chen}, \citenamefont {Lefran\c{c}ois}, \citenamefont {Vallipuram}, \citenamefont {Barth\'{e}lemy}, \citenamefont {Ataei}, \citenamefont {Yao}, \citenamefont {Li},\ and\ \citenamefont {Taillefer}}]{chen_Na2Co2TeO6_2023}%
  \BibitemOpen
  \bibfield  {author} {\bibinfo {author} {\bibfnamefont {L.}~\bibnamefont {Chen}}, \bibinfo {author} {\bibfnamefont {E.}~\bibnamefont {Lefran\c{c}ois}}, \bibinfo {author} {\bibfnamefont {A.}~\bibnamefont {Vallipuram}}, \bibinfo {author} {\bibfnamefont {Q.}~\bibnamefont {Barth\'{e}lemy}}, \bibinfo {author} {\bibfnamefont {A.}~\bibnamefont {Ataei}}, \bibinfo {author} {\bibfnamefont {W.}~\bibnamefont {Yao}}, \bibinfo {author} {\bibfnamefont {Y.}~\bibnamefont {Li}},\ and\ \bibinfo {author} {\bibfnamefont {L.}~\bibnamefont {Taillefer}},\ }\href@noop {} {} (\bibinfo {year} {2023}),\ \Eprint {https://arxiv.org/abs/2309.17231} {arXiv:2309.17231 [cond-mat.str-el]} \BibitemShut {NoStop}%
\bibitem [{\citenamefont {Hentrich}\ \emph {et~al.}(2018)\citenamefont {Hentrich}, \citenamefont {Wolter}, \citenamefont {Zotos}, \citenamefont {Brenig}, \citenamefont {Nowak}, \citenamefont {Isaeva}, \citenamefont {Doert}, \citenamefont {Banerjee}, \citenamefont {Lampen-Kelley}, \citenamefont {Mandrus}, \citenamefont {Nagler}, \citenamefont {Sears}, \citenamefont {Kim}, \citenamefont {B\"uchner},\ and\ \citenamefont {Hess}}]{Hentrich_RuCl3_2018}%
  \BibitemOpen
  \bibfield  {author} {\bibinfo {author} {\bibfnamefont {R.}~\bibnamefont {Hentrich}}, \bibinfo {author} {\bibfnamefont {A.~U.~B.}\ \bibnamefont {Wolter}}, \bibinfo {author} {\bibfnamefont {X.}~\bibnamefont {Zotos}}, \bibinfo {author} {\bibfnamefont {W.}~\bibnamefont {Brenig}}, \bibinfo {author} {\bibfnamefont {D.}~\bibnamefont {Nowak}}, \bibinfo {author} {\bibfnamefont {A.}~\bibnamefont {Isaeva}}, \bibinfo {author} {\bibfnamefont {T.}~\bibnamefont {Doert}}, \bibinfo {author} {\bibfnamefont {A.}~\bibnamefont {Banerjee}}, \bibinfo {author} {\bibfnamefont {P.}~\bibnamefont {Lampen-Kelley}}, \bibinfo {author} {\bibfnamefont {D.~G.}\ \bibnamefont {Mandrus}}, \bibinfo {author} {\bibfnamefont {S.~E.}\ \bibnamefont {Nagler}}, \bibinfo {author} {\bibfnamefont {J.}~\bibnamefont {Sears}}, \bibinfo {author} {\bibfnamefont {Y.-J.}\ \bibnamefont {Kim}}, \bibinfo {author} {\bibfnamefont {B.}~\bibnamefont {B\"uchner}},\ and\ \bibinfo {author} {\bibfnamefont {C.}~\bibnamefont {Hess}},\ }\href
  {https://doi.org/10.1103/PhysRevLett.120.117204} {\bibfield  {journal} {\bibinfo  {journal} {Phys. Rev. Lett.}\ }\textbf {\bibinfo {volume} {120}},\ \bibinfo {pages} {117204} (\bibinfo {year} {2018})}\BibitemShut {NoStop}%
\bibitem [{\citenamefont {Hentrich}\ \emph {et~al.}(2019)\citenamefont {Hentrich}, \citenamefont {Roslova}, \citenamefont {Isaeva}, \citenamefont {Doert}, \citenamefont {Brenig}, \citenamefont {B\"uchner},\ and\ \citenamefont {Hess}}]{Hentrich_RuCl3_2019}%
  \BibitemOpen
  \bibfield  {author} {\bibinfo {author} {\bibfnamefont {R.}~\bibnamefont {Hentrich}}, \bibinfo {author} {\bibfnamefont {M.}~\bibnamefont {Roslova}}, \bibinfo {author} {\bibfnamefont {A.}~\bibnamefont {Isaeva}}, \bibinfo {author} {\bibfnamefont {T.}~\bibnamefont {Doert}}, \bibinfo {author} {\bibfnamefont {W.}~\bibnamefont {Brenig}}, \bibinfo {author} {\bibfnamefont {B.}~\bibnamefont {B\"uchner}},\ and\ \bibinfo {author} {\bibfnamefont {C.}~\bibnamefont {Hess}},\ }\href {https://doi.org/10.1103/PhysRevB.99.085136} {\bibfield  {journal} {\bibinfo  {journal} {Phys. Rev. B}\ }\textbf {\bibinfo {volume} {99}},\ \bibinfo {pages} {085136} (\bibinfo {year} {2019})}\BibitemShut {NoStop}%
\bibitem [{\citenamefont {Lefran\ifmmode~\mbox{\c{c}}\else \c{c}\fi{}ois}\ \emph {et~al.}(2022)\citenamefont {Lefran\ifmmode~\mbox{\c{c}}\else \c{c}\fi{}ois}, \citenamefont {Grissonnanche}, \citenamefont {Baglo}, \citenamefont {Lampen-Kelley}, \citenamefont {Yan}, \citenamefont {Balz}, \citenamefont {Mandrus}, \citenamefont {Nagler}, \citenamefont {Kim}, \citenamefont {Kim}, \citenamefont {Doiron-Leyraud},\ and\ \citenamefont {Taillefer}}]{lefrancois_evidence_2022}%
  \BibitemOpen
  \bibfield  {author} {\bibinfo {author} {\bibfnamefont {E.}~\bibnamefont {Lefran\ifmmode~\mbox{\c{c}}\else \c{c}\fi{}ois}}, \bibinfo {author} {\bibfnamefont {G.}~\bibnamefont {Grissonnanche}}, \bibinfo {author} {\bibfnamefont {J.}~\bibnamefont {Baglo}}, \bibinfo {author} {\bibfnamefont {P.}~\bibnamefont {Lampen-Kelley}}, \bibinfo {author} {\bibfnamefont {J.-Q.}\ \bibnamefont {Yan}}, \bibinfo {author} {\bibfnamefont {C.}~\bibnamefont {Balz}}, \bibinfo {author} {\bibfnamefont {D.}~\bibnamefont {Mandrus}}, \bibinfo {author} {\bibfnamefont {S.~E.}\ \bibnamefont {Nagler}}, \bibinfo {author} {\bibfnamefont {S.}~\bibnamefont {Kim}}, \bibinfo {author} {\bibfnamefont {Y.-J.}\ \bibnamefont {Kim}}, \bibinfo {author} {\bibfnamefont {N.}~\bibnamefont {Doiron-Leyraud}},\ and\ \bibinfo {author} {\bibfnamefont {L.}~\bibnamefont {Taillefer}},\ }\href {https://doi.org/10.1103/PhysRevX.12.021025} {\bibfield  {journal} {\bibinfo  {journal} {Phys. Rev. X}\ }\textbf {\bibinfo {volume} {12}},\ \bibinfo {pages} {021025} (\bibinfo
  {year} {2022})}\BibitemShut {NoStop}%
\bibitem [{\citenamefont {Motome}\ \emph {et~al.}(2020)\citenamefont {Motome}, \citenamefont {Sano}, \citenamefont {Jang}, \citenamefont {Sugita},\ and\ \citenamefont {Kato}}]{Motome_kitaev_2020}%
  \BibitemOpen
  \bibfield  {author} {\bibinfo {author} {\bibfnamefont {Y.}~\bibnamefont {Motome}}, \bibinfo {author} {\bibfnamefont {R.}~\bibnamefont {Sano}}, \bibinfo {author} {\bibfnamefont {S.}~\bibnamefont {Jang}}, \bibinfo {author} {\bibfnamefont {Y.}~\bibnamefont {Sugita}},\ and\ \bibinfo {author} {\bibfnamefont {Y.}~\bibnamefont {Kato}},\ }\href {https://doi.org/10.1088/1361-648X/ab8525} {\bibfield  {journal} {\bibinfo  {journal} {Journal of Physics: Condensed Matter}\ }\textbf {\bibinfo {volume} {32}},\ \bibinfo {pages} {404001} (\bibinfo {year} {2020})}\BibitemShut {NoStop}%
\bibitem [{\citenamefont {Bertaut}\ \emph {et~al.}(1961)\citenamefont {Bertaut}, \citenamefont {Corliss}, \citenamefont {Forrat}, \citenamefont {Aleonard},\ and\ \citenamefont {Pauthenet}}]{Bertaut_A2B4O9_1961}%
  \BibitemOpen
  \bibfield  {author} {\bibinfo {author} {\bibfnamefont {E.}~\bibnamefont {Bertaut}}, \bibinfo {author} {\bibfnamefont {L.}~\bibnamefont {Corliss}}, \bibinfo {author} {\bibfnamefont {F.}~\bibnamefont {Forrat}}, \bibinfo {author} {\bibfnamefont {R.}~\bibnamefont {Aleonard}},\ and\ \bibinfo {author} {\bibfnamefont {R.}~\bibnamefont {Pauthenet}},\ }\href {https://doi.org/https://doi.org/10.1016/0022-3697(61)90103-2} {\bibfield  {journal} {\bibinfo  {journal} {J. Phys. Chem. Solids}\ }\textbf {\bibinfo {volume} {21}},\ \bibinfo {pages} {234} (\bibinfo {year} {1961})}\BibitemShut {NoStop}%
\bibitem [{\citenamefont {Fischer}\ \emph {et~al.}(1972)\citenamefont {Fischer}, \citenamefont {Gorodetsky},\ and\ \citenamefont {Hornreich}}]{Fischer_A2B4O9_1972}%
  \BibitemOpen
  \bibfield  {author} {\bibinfo {author} {\bibfnamefont {E.}~\bibnamefont {Fischer}}, \bibinfo {author} {\bibfnamefont {G.}~\bibnamefont {Gorodetsky}},\ and\ \bibinfo {author} {\bibfnamefont {R.}~\bibnamefont {Hornreich}},\ }\href {https://doi.org/https://doi.org/10.1016/0038-1098(72)90927-1} {\bibfield  {journal} {\bibinfo  {journal} {Solid State Commun.}\ }\textbf {\bibinfo {volume} {10}},\ \bibinfo {pages} {1127} (\bibinfo {year} {1972})}\BibitemShut {NoStop}%
\bibitem [{\citenamefont {Kolodiazhnyi}\ \emph {et~al.}(2011)\citenamefont {Kolodiazhnyi}, \citenamefont {Sakurai},\ and\ \citenamefont {Vittayakorn}}]{Kolodiazhnyi_Co4Nb2O9_2011}%
  \BibitemOpen
  \bibfield  {author} {\bibinfo {author} {\bibfnamefont {T.}~\bibnamefont {Kolodiazhnyi}}, \bibinfo {author} {\bibfnamefont {H.}~\bibnamefont {Sakurai}},\ and\ \bibinfo {author} {\bibfnamefont {N.}~\bibnamefont {Vittayakorn}},\ }\href {https://doi.org/10.1063/1.3645017} {\bibfield  {journal} {\bibinfo  {journal} {Appl. Phys. Lett.}\ }\textbf {\bibinfo {volume} {99}},\ \bibinfo {pages} {132906} (\bibinfo {year} {2011})}\BibitemShut {NoStop}%
\bibitem [{\citenamefont {Fang}\ \emph {et~al.}(2014)\citenamefont {Fang}, \citenamefont {Song}, \citenamefont {Zhou}, \citenamefont {Zhao}, \citenamefont {Tang}, \citenamefont {Yang}, \citenamefont {Lv}, \citenamefont {Yang}, \citenamefont {Wang},\ and\ \citenamefont {Du}}]{Fang_Co4Nb2O9_2014}%
  \BibitemOpen
  \bibfield  {author} {\bibinfo {author} {\bibfnamefont {Y.}~\bibnamefont {Fang}}, \bibinfo {author} {\bibfnamefont {Y.~Q.}\ \bibnamefont {Song}}, \bibinfo {author} {\bibfnamefont {W.~P.}\ \bibnamefont {Zhou}}, \bibinfo {author} {\bibfnamefont {R.}~\bibnamefont {Zhao}}, \bibinfo {author} {\bibfnamefont {R.~J.}\ \bibnamefont {Tang}}, \bibinfo {author} {\bibfnamefont {H.}~\bibnamefont {Yang}}, \bibinfo {author} {\bibfnamefont {L.~Y.}\ \bibnamefont {Lv}}, \bibinfo {author} {\bibfnamefont {S.~G.}\ \bibnamefont {Yang}}, \bibinfo {author} {\bibfnamefont {D.~H.}\ \bibnamefont {Wang}},\ and\ \bibinfo {author} {\bibfnamefont {Y.~W.}\ \bibnamefont {Du}},\ }\href {https://doi.org/10.1038/srep03860} {\bibfield  {journal} {\bibinfo  {journal} {Sci. Rep.}\ }\textbf {\bibinfo {volume} {4}},\ \bibinfo {pages} {3860} (\bibinfo {year} {2014})}\BibitemShut {NoStop}%
\bibitem [{\citenamefont {Fang}\ \emph {et~al.}(2015)\citenamefont {Fang}, \citenamefont {Yan}, \citenamefont {Zhang}, \citenamefont {Han}, \citenamefont {Qian}, \citenamefont {Wang},\ and\ \citenamefont {Du}}]{fang_Co4Ta2O9_2015}%
  \BibitemOpen
  \bibfield  {author} {\bibinfo {author} {\bibfnamefont {Y.}~\bibnamefont {Fang}}, \bibinfo {author} {\bibfnamefont {S.}~\bibnamefont {Yan}}, \bibinfo {author} {\bibfnamefont {L.}~\bibnamefont {Zhang}}, \bibinfo {author} {\bibfnamefont {Z.}~\bibnamefont {Han}}, \bibinfo {author} {\bibfnamefont {B.}~\bibnamefont {Qian}}, \bibinfo {author} {\bibfnamefont {D.}~\bibnamefont {Wang}},\ and\ \bibinfo {author} {\bibfnamefont {Y.}~\bibnamefont {Du}},\ }\href {https://doi.org/10.1111/jace.13651} {\bibfield  {journal} {\bibinfo  {journal} {J. Am. Ceram. Soc.}\ }\textbf {\bibinfo {volume} {98}},\ \bibinfo {pages} {2005} (\bibinfo {year} {2015})}\BibitemShut {NoStop}%
\bibitem [{\citenamefont {Khanh}\ \emph {et~al.}(2016)\citenamefont {Khanh}, \citenamefont {Abe}, \citenamefont {Sagayama}, \citenamefont {Nakao}, \citenamefont {Hanashima}, \citenamefont {Kiyanagi}, \citenamefont {Tokunaga},\ and\ \citenamefont {Arima}}]{khanh_Co4Nb2O9_2016}%
  \BibitemOpen
  \bibfield  {author} {\bibinfo {author} {\bibfnamefont {N.~D.}\ \bibnamefont {Khanh}}, \bibinfo {author} {\bibfnamefont {N.}~\bibnamefont {Abe}}, \bibinfo {author} {\bibfnamefont {H.}~\bibnamefont {Sagayama}}, \bibinfo {author} {\bibfnamefont {A.}~\bibnamefont {Nakao}}, \bibinfo {author} {\bibfnamefont {T.}~\bibnamefont {Hanashima}}, \bibinfo {author} {\bibfnamefont {R.}~\bibnamefont {Kiyanagi}}, \bibinfo {author} {\bibfnamefont {Y.}~\bibnamefont {Tokunaga}},\ and\ \bibinfo {author} {\bibfnamefont {T.}~\bibnamefont {Arima}},\ }\href {https://doi.org/10.1103/PhysRevB.93.075117} {\bibfield  {journal} {\bibinfo  {journal} {Phys. Rev. B}\ }\textbf {\bibinfo {volume} {93}},\ \bibinfo {pages} {075117} (\bibinfo {year} {2016})}\BibitemShut {NoStop}%
\bibitem [{\citenamefont {Xie}\ \emph {et~al.}(2016)\citenamefont {Xie}, \citenamefont {Lin}, \citenamefont {Zhang},\ and\ \citenamefont {Cheng}}]{Xie_Co4Nb2O9_2016}%
  \BibitemOpen
  \bibfield  {author} {\bibinfo {author} {\bibfnamefont {Y.~M.}\ \bibnamefont {Xie}}, \bibinfo {author} {\bibfnamefont {C.~S.}\ \bibnamefont {Lin}}, \bibinfo {author} {\bibfnamefont {H.}~\bibnamefont {Zhang}},\ and\ \bibinfo {author} {\bibfnamefont {W.~D.}\ \bibnamefont {Cheng}},\ }\href {https://doi.org/10.1063/1.4947074} {\bibfield  {journal} {\bibinfo  {journal} {AIP Adv.}\ }\textbf {\bibinfo {volume} {6}},\ \bibinfo {pages} {045006} (\bibinfo {year} {2016})}\BibitemShut {NoStop}%
\bibitem [{\citenamefont {Yin}\ \emph {et~al.}(2016)\citenamefont {Yin}, \citenamefont {Zou}, \citenamefont {Yang}, \citenamefont {Dai}, \citenamefont {Song}, \citenamefont {Zhu},\ and\ \citenamefont {Sun}}]{Yin_Co4Nb2O9_2016}%
  \BibitemOpen
  \bibfield  {author} {\bibinfo {author} {\bibfnamefont {L.~H.}\ \bibnamefont {Yin}}, \bibinfo {author} {\bibfnamefont {Y.~M.}\ \bibnamefont {Zou}}, \bibinfo {author} {\bibfnamefont {J.}~\bibnamefont {Yang}}, \bibinfo {author} {\bibfnamefont {J.~M.}\ \bibnamefont {Dai}}, \bibinfo {author} {\bibfnamefont {W.~H.}\ \bibnamefont {Song}}, \bibinfo {author} {\bibfnamefont {X.~B.}\ \bibnamefont {Zhu}},\ and\ \bibinfo {author} {\bibfnamefont {Y.~P.}\ \bibnamefont {Sun}},\ }\href {https://doi.org/10.1063/1.4959086} {\bibfield  {journal} {\bibinfo  {journal} {Appl. Phys. Lett.}\ }\textbf {\bibinfo {volume} {109}},\ \bibinfo {pages} {032905} (\bibinfo {year} {2016})}\BibitemShut {NoStop}%
\bibitem [{\citenamefont {Lu}\ \emph {et~al.}(2016)\citenamefont {Lu}, \citenamefont {Ji}, \citenamefont {Sun}, \citenamefont {Fang}, \citenamefont {Zhang}, \citenamefont {Han}, \citenamefont {Qian}, \citenamefont {Jiang},\ and\ \citenamefont {Zhou}}]{Lu_Co4Nb2O9_2016}%
  \BibitemOpen
  \bibfield  {author} {\bibinfo {author} {\bibfnamefont {Y.}~\bibnamefont {Lu}}, \bibinfo {author} {\bibfnamefont {C.}~\bibnamefont {Ji}}, \bibinfo {author} {\bibfnamefont {Y.}~\bibnamefont {Sun}}, \bibinfo {author} {\bibfnamefont {Y.}~\bibnamefont {Fang}}, \bibinfo {author} {\bibfnamefont {L.}~\bibnamefont {Zhang}}, \bibinfo {author} {\bibfnamefont {Z.}~\bibnamefont {Han}}, \bibinfo {author} {\bibfnamefont {B.}~\bibnamefont {Qian}}, \bibinfo {author} {\bibfnamefont {X.}~\bibnamefont {Jiang}},\ and\ \bibinfo {author} {\bibfnamefont {W.}~\bibnamefont {Zhou}},\ }\href {https://doi.org/https://doi.org/10.1016/j.jallcom.2016.04.053} {\bibfield  {journal} {\bibinfo  {journal} {J. Alloys Compd.}\ }\textbf {\bibinfo {volume} {679}},\ \bibinfo {pages} {213} (\bibinfo {year} {2016})}\BibitemShut {NoStop}%
\bibitem [{\citenamefont {Khanh}\ \emph {et~al.}(2017)\citenamefont {Khanh}, \citenamefont {Abe}, \citenamefont {Kimura}, \citenamefont {Tokunaga},\ and\ \citenamefont {Arima}}]{khanh_manipulation_2017}%
  \BibitemOpen
  \bibfield  {author} {\bibinfo {author} {\bibfnamefont {N.~D.}\ \bibnamefont {Khanh}}, \bibinfo {author} {\bibfnamefont {N.}~\bibnamefont {Abe}}, \bibinfo {author} {\bibfnamefont {S.}~\bibnamefont {Kimura}}, \bibinfo {author} {\bibfnamefont {Y.}~\bibnamefont {Tokunaga}},\ and\ \bibinfo {author} {\bibfnamefont {T.}~\bibnamefont {Arima}},\ }\href {https://doi.org/10.1103/PhysRevB.96.094434} {\bibfield  {journal} {\bibinfo  {journal} {Phys. Rev. B}\ }\textbf {\bibinfo {volume} {96}},\ \bibinfo {pages} {094434} (\bibinfo {year} {2017})}\BibitemShut {NoStop}%
\bibitem [{\citenamefont {Cao}\ \emph {et~al.}(2017)\citenamefont {Cao}, \citenamefont {Deng}, \citenamefont {Beran}, \citenamefont {Feng}, \citenamefont {Kang}, \citenamefont {Zhang}, \citenamefont {Guiblin}, \citenamefont {Dkhil}, \citenamefont {Ren},\ and\ \citenamefont {Cao}}]{cao_Co4Nb2O9_2017}%
  \BibitemOpen
  \bibfield  {author} {\bibinfo {author} {\bibfnamefont {Y.}~\bibnamefont {Cao}}, \bibinfo {author} {\bibfnamefont {G.}~\bibnamefont {Deng}}, \bibinfo {author} {\bibfnamefont {P.}~\bibnamefont {Beran}}, \bibinfo {author} {\bibfnamefont {Z.}~\bibnamefont {Feng}}, \bibinfo {author} {\bibfnamefont {B.}~\bibnamefont {Kang}}, \bibinfo {author} {\bibfnamefont {J.}~\bibnamefont {Zhang}}, \bibinfo {author} {\bibfnamefont {N.}~\bibnamefont {Guiblin}}, \bibinfo {author} {\bibfnamefont {B.}~\bibnamefont {Dkhil}}, \bibinfo {author} {\bibfnamefont {W.}~\bibnamefont {Ren}},\ and\ \bibinfo {author} {\bibfnamefont {S.}~\bibnamefont {Cao}},\ }\href {https://doi.org/10.1038/s41598-017-14169-3} {\bibfield  {journal} {\bibinfo  {journal} {Sci. Rep.}\ }\textbf {\bibinfo {volume} {7}},\ \bibinfo {pages} {14079} (\bibinfo {year} {2017})}\BibitemShut {NoStop}%
\bibitem [{\citenamefont {Chaudhary}\ \emph {et~al.}(2019)\citenamefont {Chaudhary}, \citenamefont {Srivastava}, \citenamefont {Kaushik}, \citenamefont {Siruguri},\ and\ \citenamefont {Patnaik}}]{chaudhary_nature_2019}%
  \BibitemOpen
  \bibfield  {author} {\bibinfo {author} {\bibfnamefont {S.}~\bibnamefont {Chaudhary}}, \bibinfo {author} {\bibfnamefont {P.}~\bibnamefont {Srivastava}}, \bibinfo {author} {\bibfnamefont {S.}~\bibnamefont {Kaushik}}, \bibinfo {author} {\bibfnamefont {V.}~\bibnamefont {Siruguri}},\ and\ \bibinfo {author} {\bibfnamefont {S.}~\bibnamefont {Patnaik}},\ }\href {https://doi.org/10.1016/j.jmmm.2018.09.071} {\bibfield  {journal} {\bibinfo  {journal} {J. Magn. Magn. Mater.}\ }\textbf {\bibinfo {volume} {475}},\ \bibinfo {pages} {508} (\bibinfo {year} {2019})}\BibitemShut {NoStop}%
\bibitem [{\citenamefont {Lee}\ \emph {et~al.}(2020)\citenamefont {Lee}, \citenamefont {Oh}, \citenamefont {Choi}, \citenamefont {Moon}, \citenamefont {Kim}, \citenamefont {Shin}, \citenamefont {Son}, \citenamefont {Nuss}, \citenamefont {Kiryukhin},\ and\ \citenamefont {Choi}}]{lee_highly_2020}%
  \BibitemOpen
  \bibfield  {author} {\bibinfo {author} {\bibfnamefont {N.}~\bibnamefont {Lee}}, \bibinfo {author} {\bibfnamefont {D.~G.}\ \bibnamefont {Oh}}, \bibinfo {author} {\bibfnamefont {S.}~\bibnamefont {Choi}}, \bibinfo {author} {\bibfnamefont {J.~Y.}\ \bibnamefont {Moon}}, \bibinfo {author} {\bibfnamefont {J.~H.}\ \bibnamefont {Kim}}, \bibinfo {author} {\bibfnamefont {H.~J.}\ \bibnamefont {Shin}}, \bibinfo {author} {\bibfnamefont {K.}~\bibnamefont {Son}}, \bibinfo {author} {\bibfnamefont {J.}~\bibnamefont {Nuss}}, \bibinfo {author} {\bibfnamefont {V.}~\bibnamefont {Kiryukhin}},\ and\ \bibinfo {author} {\bibfnamefont {Y.~J.}\ \bibnamefont {Choi}},\ }\href {https://doi.org/10.1038/s41598-020-69117-5} {\bibfield  {journal} {\bibinfo  {journal} {Sci. Rep.}\ }\textbf {\bibinfo {volume} {10}},\ \bibinfo {pages} {12362} (\bibinfo {year} {2020})}\BibitemShut {NoStop}%
\bibitem [{\citenamefont {Choi}\ \emph {et~al.}(2020)\citenamefont {Choi}, \citenamefont {Oh}, \citenamefont {Gutmann}, \citenamefont {Pan}, \citenamefont {Kim}, \citenamefont {Son}, \citenamefont {Kim}, \citenamefont {Lee}, \citenamefont {Cheong}, \citenamefont {Choi},\ and\ \citenamefont {Kiryukhin}}]{choi_noncollinear_2020}%
  \BibitemOpen
  \bibfield  {author} {\bibinfo {author} {\bibfnamefont {S.}~\bibnamefont {Choi}}, \bibinfo {author} {\bibfnamefont {D.~G.}\ \bibnamefont {Oh}}, \bibinfo {author} {\bibfnamefont {M.~J.}\ \bibnamefont {Gutmann}}, \bibinfo {author} {\bibfnamefont {S.}~\bibnamefont {Pan}}, \bibinfo {author} {\bibfnamefont {G.}~\bibnamefont {Kim}}, \bibinfo {author} {\bibfnamefont {K.}~\bibnamefont {Son}}, \bibinfo {author} {\bibfnamefont {J.}~\bibnamefont {Kim}}, \bibinfo {author} {\bibfnamefont {N.}~\bibnamefont {Lee}}, \bibinfo {author} {\bibfnamefont {S.-W.}\ \bibnamefont {Cheong}}, \bibinfo {author} {\bibfnamefont {Y.~J.}\ \bibnamefont {Choi}},\ and\ \bibinfo {author} {\bibfnamefont {V.}~\bibnamefont {Kiryukhin}},\ }\href {https://doi.org/10.1103/PhysRevB.102.214404} {\bibfield  {journal} {\bibinfo  {journal} {Phys. Rev. B}\ }\textbf {\bibinfo {volume} {102}},\ \bibinfo {pages} {214404} (\bibinfo {year} {2020})}\BibitemShut {NoStop}%
\bibitem [{\citenamefont {Deng}\ \emph {et~al.}(2018)\citenamefont {Deng}, \citenamefont {Cao}, \citenamefont {Ren}, \citenamefont {Cao}, \citenamefont {Studer}, \citenamefont {Gauthier}, \citenamefont {Kenzelmann}, \citenamefont {Davidson}, \citenamefont {Rule}, \citenamefont {Gardner}, \citenamefont {Imperia}, \citenamefont {Ulrich},\ and\ \citenamefont {{McIntyre}}}]{deng_spin_2018}%
  \BibitemOpen
  \bibfield  {author} {\bibinfo {author} {\bibfnamefont {G.}~\bibnamefont {Deng}}, \bibinfo {author} {\bibfnamefont {Y.}~\bibnamefont {Cao}}, \bibinfo {author} {\bibfnamefont {W.}~\bibnamefont {Ren}}, \bibinfo {author} {\bibfnamefont {S.}~\bibnamefont {Cao}}, \bibinfo {author} {\bibfnamefont {A.~J.}\ \bibnamefont {Studer}}, \bibinfo {author} {\bibfnamefont {N.}~\bibnamefont {Gauthier}}, \bibinfo {author} {\bibfnamefont {M.}~\bibnamefont {Kenzelmann}}, \bibinfo {author} {\bibfnamefont {G.}~\bibnamefont {Davidson}}, \bibinfo {author} {\bibfnamefont {K.~C.}\ \bibnamefont {Rule}}, \bibinfo {author} {\bibfnamefont {J.~S.}\ \bibnamefont {Gardner}}, \bibinfo {author} {\bibfnamefont {P.}~\bibnamefont {Imperia}}, \bibinfo {author} {\bibfnamefont {C.}~\bibnamefont {Ulrich}},\ and\ \bibinfo {author} {\bibfnamefont {G.~J.}\ \bibnamefont {{McIntyre}}},\ }\href {https://doi.org/10.1103/PhysRevB.97.085154} {\bibfield  {journal} {\bibinfo  {journal} {Phys. Rev. B}\ }\textbf {\bibinfo {volume} {97}},\ \bibinfo {pages}
  {085154} (\bibinfo {year} {2018})}\BibitemShut {NoStop}%
\bibitem [{\citenamefont {Ding}\ \emph {et~al.}(2020)\citenamefont {Ding}, \citenamefont {Lee}, \citenamefont {Hong}, \citenamefont {Dun}, \citenamefont {Sinclair}, \citenamefont {Chi}, \citenamefont {Agrawal}, \citenamefont {Choi}, \citenamefont {Chakoumakos}, \citenamefont {Zhou},\ and\ \citenamefont {Cao}}]{ding_Co4Nb2O9_2020}%
  \BibitemOpen
  \bibfield  {author} {\bibinfo {author} {\bibfnamefont {L.}~\bibnamefont {Ding}}, \bibinfo {author} {\bibfnamefont {M.}~\bibnamefont {Lee}}, \bibinfo {author} {\bibfnamefont {T.}~\bibnamefont {Hong}}, \bibinfo {author} {\bibfnamefont {Z.}~\bibnamefont {Dun}}, \bibinfo {author} {\bibfnamefont {R.}~\bibnamefont {Sinclair}}, \bibinfo {author} {\bibfnamefont {S.}~\bibnamefont {Chi}}, \bibinfo {author} {\bibfnamefont {H.~K.}\ \bibnamefont {Agrawal}}, \bibinfo {author} {\bibfnamefont {E.~S.}\ \bibnamefont {Choi}}, \bibinfo {author} {\bibfnamefont {B.~C.}\ \bibnamefont {Chakoumakos}}, \bibinfo {author} {\bibfnamefont {H.}~\bibnamefont {Zhou}},\ and\ \bibinfo {author} {\bibfnamefont {H.}~\bibnamefont {Cao}},\ }\href {https://doi.org/10.1103/PhysRevB.102.174443} {\bibfield  {journal} {\bibinfo  {journal} {Phys. Rev. B}\ }\textbf {\bibinfo {volume} {102}},\ \bibinfo {pages} {174443} (\bibinfo {year} {2020})}\BibitemShut {NoStop}%
\bibitem [{\citenamefont {Cao}\ \emph {et~al.}(2015)\citenamefont {Cao}, \citenamefont {Yang}, \citenamefont {Xiang}, \citenamefont {Feng}, \citenamefont {Kang}, \citenamefont {Zhang}, \citenamefont {Ren},\ and\ \citenamefont {Cao}}]{Cao_Co4Nb2O9_2015}%
  \BibitemOpen
  \bibfield  {author} {\bibinfo {author} {\bibfnamefont {Y.}~\bibnamefont {Cao}}, \bibinfo {author} {\bibfnamefont {Y.}~\bibnamefont {Yang}}, \bibinfo {author} {\bibfnamefont {M.}~\bibnamefont {Xiang}}, \bibinfo {author} {\bibfnamefont {Z.}~\bibnamefont {Feng}}, \bibinfo {author} {\bibfnamefont {B.}~\bibnamefont {Kang}}, \bibinfo {author} {\bibfnamefont {J.}~\bibnamefont {Zhang}}, \bibinfo {author} {\bibfnamefont {W.}~\bibnamefont {Ren}},\ and\ \bibinfo {author} {\bibfnamefont {S.}~\bibnamefont {Cao}},\ }\href {https://doi.org/https://doi.org/10.1016/j.jcrysgro.2015.03.045} {\bibfield  {journal} {\bibinfo  {journal} {J. Cryst. Growth}\ }\textbf {\bibinfo {volume} {420}},\ \bibinfo {pages} {90} (\bibinfo {year} {2015})}\BibitemShut {NoStop}%
\bibitem [{\citenamefont {Momma}\ and\ \citenamefont {Izumi}(2011)}]{momma_vesta_2011}%
  \BibitemOpen
  \bibfield  {author} {\bibinfo {author} {\bibfnamefont {K.}~\bibnamefont {Momma}}\ and\ \bibinfo {author} {\bibfnamefont {F.}~\bibnamefont {Izumi}},\ }\href {https://doi.org/10.1107/S0021889811038970} {\bibfield  {journal} {\bibinfo  {journal} {J. Appl. Cryst.}\ }\textbf {\bibinfo {volume} {44}},\ \bibinfo {pages} {1272} (\bibinfo {year} {2011})}\BibitemShut {NoStop}%
\bibitem [{Uen()}]{Ueno_SupplemntalMaterial_2023}%
  \BibitemOpen
  \href@noop {} {\bibfield  {journal} {\bibinfo  {journal} {See Supplemental Material at [url] for an extended discussion and details of the derivation, including Refs. [83-91], which are quoted therein}\ }\textbf {\bibinfo {volume} {{}}}}\BibitemShut {NoStop}%
\bibitem [{\citenamefont {Gen}\ \emph {et~al.}(2022)\citenamefont {Gen}, \citenamefont {Miyake}, \citenamefont {Yagiuchi}, \citenamefont {Watanabe}, \citenamefont {Ikeda}, \citenamefont {Matsuda}, \citenamefont {Tokunaga}, \citenamefont {Arima},\ and\ \citenamefont {Tokunaga}}]{Gen_FBG_2022}%
  \BibitemOpen
  \bibfield  {author} {\bibinfo {author} {\bibfnamefont {M.}~\bibnamefont {Gen}}, \bibinfo {author} {\bibfnamefont {A.}~\bibnamefont {Miyake}}, \bibinfo {author} {\bibfnamefont {H.}~\bibnamefont {Yagiuchi}}, \bibinfo {author} {\bibfnamefont {Y.}~\bibnamefont {Watanabe}}, \bibinfo {author} {\bibfnamefont {A.}~\bibnamefont {Ikeda}}, \bibinfo {author} {\bibfnamefont {Y.~H.}\ \bibnamefont {Matsuda}}, \bibinfo {author} {\bibfnamefont {M.}~\bibnamefont {Tokunaga}}, \bibinfo {author} {\bibfnamefont {T.}~\bibnamefont {Arima}},\ and\ \bibinfo {author} {\bibfnamefont {Y.}~\bibnamefont {Tokunaga}},\ }\href {https://doi.org/10.1103/PhysRevB.105.214412} {\bibfield  {journal} {\bibinfo  {journal} {Phys. Rev. B}\ }\textbf {\bibinfo {volume} {105}},\ \bibinfo {pages} {214412} (\bibinfo {year} {2022})}\BibitemShut {NoStop}%
\bibitem [{\citenamefont {Slack}(1961)}]{Slack_Mn_1961}%
  \BibitemOpen
  \bibfield  {author} {\bibinfo {author} {\bibfnamefont {G.~A.}\ \bibnamefont {Slack}},\ }\href {https://doi.org/10.1103/PhysRev.122.1451} {\bibfield  {journal} {\bibinfo  {journal} {Phys. Rev.}\ }\textbf {\bibinfo {volume} {122}},\ \bibinfo {pages} {1451} (\bibinfo {year} {1961})}\BibitemShut {NoStop}%
\bibitem [{\citenamefont {Ozhogin}\ \emph {et~al.}(1983)\citenamefont {Ozhogin}, \citenamefont {Inyushkin},\ and\ \citenamefont {Babushkina}}]{Ozhogin_CoCO3MnCO3_1983}%
  \BibitemOpen
  \bibfield  {author} {\bibinfo {author} {\bibfnamefont {V.}~\bibnamefont {Ozhogin}}, \bibinfo {author} {\bibfnamefont {A.}~\bibnamefont {Inyushkin}},\ and\ \bibinfo {author} {\bibfnamefont {N.}~\bibnamefont {Babushkina}},\ }\href {https://doi.org/https://doi.org/10.1016/0304-8853(83)90191-9} {\bibfield  {journal} {\bibinfo  {journal} {J. Magn. Magn. Mater.}\ }\textbf {\bibinfo {volume} {31-34}},\ \bibinfo {pages} {147} (\bibinfo {year} {1983})}\BibitemShut {NoStop}%
\bibitem [{\citenamefont {Rives}\ \emph {et~al.}(1969)\citenamefont {Rives}, \citenamefont {Dixon},\ and\ \citenamefont {Walton}}]{rives1969effect}%
  \BibitemOpen
  \bibfield  {author} {\bibinfo {author} {\bibfnamefont {J.~E.}\ \bibnamefont {Rives}}, \bibinfo {author} {\bibfnamefont {G.~S.}\ \bibnamefont {Dixon}},\ and\ \bibinfo {author} {\bibfnamefont {D.}~\bibnamefont {Walton}},\ }\href@noop {} {\bibfield  {journal} {\bibinfo  {journal} {J. Appl. Phys.}\ }\textbf {\bibinfo {volume} {40}},\ \bibinfo {pages} {1555} (\bibinfo {year} {1969})}\BibitemShut {NoStop}%
\bibitem [{\citenamefont {Gustafson}\ and\ \citenamefont {Walker}(1973)}]{gustafson1973thermal}%
  \BibitemOpen
  \bibfield  {author} {\bibinfo {author} {\bibfnamefont {J.}~\bibnamefont {Gustafson}}\ and\ \bibinfo {author} {\bibfnamefont {C.~T.}\ \bibnamefont {Walker}},\ }\href@noop {} {\bibfield  {journal} {\bibinfo  {journal} {Phys. Rev. B}\ }\textbf {\bibinfo {volume} {8}},\ \bibinfo {pages} {3309} (\bibinfo {year} {1973})}\BibitemShut {NoStop}%
\bibitem [{\citenamefont {Sanders}\ and\ \citenamefont {Walton}(1977)}]{sanders_effect_1977}%
  \BibitemOpen
  \bibfield  {author} {\bibinfo {author} {\bibfnamefont {D.~J.}\ \bibnamefont {Sanders}}\ and\ \bibinfo {author} {\bibfnamefont {D.}~\bibnamefont {Walton}},\ }\href {https://doi.org/10.1103/PhysRevB.15.1489} {\bibfield  {journal} {\bibinfo  {journal} {Phys. Rev. B}\ }\textbf {\bibinfo {volume} {15}},\ \bibinfo {pages} {1489} (\bibinfo {year} {1977})}\BibitemShut {NoStop}%
\bibitem [{\citenamefont {Callaway}(1959)}]{Callaway_CallawayModel_1959}%
  \BibitemOpen
  \bibfield  {author} {\bibinfo {author} {\bibfnamefont {J.}~\bibnamefont {Callaway}},\ }\href {https://doi.org/10.1103/PhysRev.113.1046} {\bibfield  {journal} {\bibinfo  {journal} {Phys. Rev.}\ }\textbf {\bibinfo {volume} {113}},\ \bibinfo {pages} {1046} (\bibinfo {year} {1959})}\BibitemShut {NoStop}%
\bibitem [{\citenamefont {Callaway}(1961)}]{Callaway_Callaway_1961}%
  \BibitemOpen
  \bibfield  {author} {\bibinfo {author} {\bibfnamefont {J.}~\bibnamefont {Callaway}},\ }\href {https://doi.org/10.1103/PhysRev.122.787} {\bibfield  {journal} {\bibinfo  {journal} {Phys. Rev.}\ }\textbf {\bibinfo {volume} {122}},\ \bibinfo {pages} {787} (\bibinfo {year} {1961})}\BibitemShut {NoStop}%
\bibitem [{\citenamefont {Sheard}\ and\ \citenamefont {Toombs}(1973)}]{SHEARD_phonon_1973}%
  \BibitemOpen
  \bibfield  {author} {\bibinfo {author} {\bibfnamefont {F.}~\bibnamefont {Sheard}}\ and\ \bibinfo {author} {\bibfnamefont {G.}~\bibnamefont {Toombs}},\ }\href {https://doi.org/https://doi.org/10.1016/0038-1098(73)90320-7} {\bibfield  {journal} {\bibinfo  {journal} {Solid State Commun.}\ }\textbf {\bibinfo {volume} {12}},\ \bibinfo {pages} {713} (\bibinfo {year} {1973})}\BibitemShut {NoStop}%
\bibitem [{\citenamefont {Wybourne}\ and\ \citenamefont {Kiff}(1985)}]{Wybourne_phonon_1985}%
  \BibitemOpen
  \bibfield  {author} {\bibinfo {author} {\bibfnamefont {M.~N.}\ \bibnamefont {Wybourne}}\ and\ \bibinfo {author} {\bibfnamefont {B.~J.}\ \bibnamefont {Kiff}},\ }\href {https://doi.org/10.1088/0022-3719/18/2/010} {\bibfield  {journal} {\bibinfo  {journal} {J. Phys. C}\ }\textbf {\bibinfo {volume} {18}},\ \bibinfo {pages} {309} (\bibinfo {year} {1985})}\BibitemShut {NoStop}%
\bibitem [{\citenamefont {Prasai}\ \emph {et~al.}(2018)\citenamefont {Prasai}, \citenamefont {Christian}, \citenamefont {Neumeier},\ and\ \citenamefont {Cohn}}]{Prasai_AB2O6_2018}%
  \BibitemOpen
  \bibfield  {author} {\bibinfo {author} {\bibfnamefont {N.}~\bibnamefont {Prasai}}, \bibinfo {author} {\bibfnamefont {A.~B.}\ \bibnamefont {Christian}}, \bibinfo {author} {\bibfnamefont {J.~J.}\ \bibnamefont {Neumeier}},\ and\ \bibinfo {author} {\bibfnamefont {J.~L.}\ \bibnamefont {Cohn}},\ }\href {https://doi.org/10.1103/PhysRevB.98.134449} {\bibfield  {journal} {\bibinfo  {journal} {Phys. Rev. B}\ }\textbf {\bibinfo {volume} {98}},\ \bibinfo {pages} {134449} (\bibinfo {year} {2018})}\BibitemShut {NoStop}%
\bibitem [{\citenamefont {Hentrich}\ \emph {et~al.}(2020)\citenamefont {Hentrich}, \citenamefont {Hong}, \citenamefont {Gillig}, \citenamefont {Caglieris}, \citenamefont {\ifmmode~\check{C}\else \v{C}\fi{}ulo}, \citenamefont {Shahrokhvand}, \citenamefont {Zeitler}, \citenamefont {Roslova}, \citenamefont {Isaeva}, \citenamefont {Doert}, \citenamefont {Janssen}, \citenamefont {Vojta}, \citenamefont {B\"uchner},\ and\ \citenamefont {Hess}}]{Hentrich_RuCl3_2020}%
  \BibitemOpen
  \bibfield  {author} {\bibinfo {author} {\bibfnamefont {R.}~\bibnamefont {Hentrich}}, \bibinfo {author} {\bibfnamefont {X.}~\bibnamefont {Hong}}, \bibinfo {author} {\bibfnamefont {M.}~\bibnamefont {Gillig}}, \bibinfo {author} {\bibfnamefont {F.}~\bibnamefont {Caglieris}}, \bibinfo {author} {\bibfnamefont {M.}~\bibnamefont {\ifmmode~\check{C}\else \v{C}\fi{}ulo}}, \bibinfo {author} {\bibfnamefont {M.}~\bibnamefont {Shahrokhvand}}, \bibinfo {author} {\bibfnamefont {U.}~\bibnamefont {Zeitler}}, \bibinfo {author} {\bibfnamefont {M.}~\bibnamefont {Roslova}}, \bibinfo {author} {\bibfnamefont {A.}~\bibnamefont {Isaeva}}, \bibinfo {author} {\bibfnamefont {T.}~\bibnamefont {Doert}}, \bibinfo {author} {\bibfnamefont {L.}~\bibnamefont {Janssen}}, \bibinfo {author} {\bibfnamefont {M.}~\bibnamefont {Vojta}}, \bibinfo {author} {\bibfnamefont {B.}~\bibnamefont {B\"uchner}},\ and\ \bibinfo {author} {\bibfnamefont {C.}~\bibnamefont {Hess}},\ }\href {https://doi.org/10.1103/PhysRevB.102.235155} {\bibfield  {journal} {\bibinfo
   {journal} {Phys. Rev. B}\ }\textbf {\bibinfo {volume} {102}},\ \bibinfo {pages} {235155} (\bibinfo {year} {2020})}\BibitemShut {NoStop}%
\bibitem [{\citenamefont {Yang}\ \emph {et~al.}(2023)\citenamefont {Yang}, \citenamefont {Wu}, \citenamefont {Li}, \citenamefont {Ran}, \citenamefont {Wang}, \citenamefont {Zhu}, \citenamefont {Gong}, \citenamefont {Liu}, \citenamefont {Wang}, \citenamefont {Zhang}, \citenamefont {Mi}, \citenamefont {Wang}, \citenamefont {Chai}, \citenamefont {Su}, \citenamefont {Wang}, \citenamefont {He}, \citenamefont {Yang},\ and\ \citenamefont {Zhou}}]{yang_Cr2Si2Te6_2023}%
  \BibitemOpen
  \bibfield  {author} {\bibinfo {author} {\bibfnamefont {K.}~\bibnamefont {Yang}}, \bibinfo {author} {\bibfnamefont {H.}~\bibnamefont {Wu}}, \bibinfo {author} {\bibfnamefont {Z.}~\bibnamefont {Li}}, \bibinfo {author} {\bibfnamefont {C.}~\bibnamefont {Ran}}, \bibinfo {author} {\bibfnamefont {X.}~\bibnamefont {Wang}}, \bibinfo {author} {\bibfnamefont {F.}~\bibnamefont {Zhu}}, \bibinfo {author} {\bibfnamefont {X.}~\bibnamefont {Gong}}, \bibinfo {author} {\bibfnamefont {Y.}~\bibnamefont {Liu}}, \bibinfo {author} {\bibfnamefont {G.}~\bibnamefont {Wang}}, \bibinfo {author} {\bibfnamefont {L.}~\bibnamefont {Zhang}}, \bibinfo {author} {\bibfnamefont {X.}~\bibnamefont {Mi}}, \bibinfo {author} {\bibfnamefont {A.}~\bibnamefont {Wang}}, \bibinfo {author} {\bibfnamefont {Y.}~\bibnamefont {Chai}}, \bibinfo {author} {\bibfnamefont {Y.}~\bibnamefont {Su}}, \bibinfo {author} {\bibfnamefont {W.}~\bibnamefont {Wang}}, \bibinfo {author} {\bibfnamefont {M.}~\bibnamefont {He}}, \bibinfo {author} {\bibfnamefont {X.}~\bibnamefont
  {Yang}},\ and\ \bibinfo {author} {\bibfnamefont {X.}~\bibnamefont {Zhou}},\ }\href {https://doi.org/https://doi.org/10.1002/adfm.202302191} {\bibfield  {journal} {\bibinfo  {journal} {Adv. Funct. Mater.}\ }\textbf {\bibinfo {volume} {33}},\ \bibinfo {pages} {2302191} (\bibinfo {year} {2023})}\BibitemShut {NoStop}%
\bibitem [{\citenamefont {Park}\ \emph {et~al.}(2023)\citenamefont {Park}, \citenamefont {Kim}, \citenamefont {Choi}, \citenamefont {Fan}, \citenamefont {Kim}, \citenamefont {Oh}, \citenamefont {Lee}, \citenamefont {Cheong}, \citenamefont {Kiryukhin}, \citenamefont {Choi}, \citenamefont {Vanderbilt}, \citenamefont {Lee},\ and\ \citenamefont {Musfeldt}}]{park_Co4Nb2O9-Co4Ta2O9_2023}%
  \BibitemOpen
  \bibfield  {author} {\bibinfo {author} {\bibfnamefont {K.}~\bibnamefont {Park}}, \bibinfo {author} {\bibfnamefont {J.}~\bibnamefont {Kim}}, \bibinfo {author} {\bibfnamefont {S.}~\bibnamefont {Choi}}, \bibinfo {author} {\bibfnamefont {S.}~\bibnamefont {Fan}}, \bibinfo {author} {\bibfnamefont {C.}~\bibnamefont {Kim}}, \bibinfo {author} {\bibfnamefont {D.~G.}\ \bibnamefont {Oh}}, \bibinfo {author} {\bibfnamefont {N.}~\bibnamefont {Lee}}, \bibinfo {author} {\bibfnamefont {S.-W.}\ \bibnamefont {Cheong}}, \bibinfo {author} {\bibfnamefont {V.}~\bibnamefont {Kiryukhin}}, \bibinfo {author} {\bibfnamefont {Y.~J.}\ \bibnamefont {Choi}}, \bibinfo {author} {\bibfnamefont {D.}~\bibnamefont {Vanderbilt}}, \bibinfo {author} {\bibfnamefont {J.~H.}\ \bibnamefont {Lee}},\ and\ \bibinfo {author} {\bibfnamefont {J.~L.}\ \bibnamefont {Musfeldt}},\ }\href {https://doi.org/10.1063/5.0137903} {\bibfield  {journal} {\bibinfo  {journal} {Appl. Phys. Lett.}\ }\textbf {\bibinfo {volume} {122}},\ \bibinfo {pages} {182902} (\bibinfo
  {year} {2023})}\BibitemShut {NoStop}%
\bibitem [{\citenamefont {Dagar}\ \emph {et~al.}(2022)\citenamefont {Dagar}, \citenamefont {Yadav}, \citenamefont {Kinha}, \citenamefont {Mehra}, \citenamefont {Rawat}, \citenamefont {Singh},\ and\ \citenamefont {Rana}}]{dagar_Co4Nb2O9_2022}%
  \BibitemOpen
  \bibfield  {author} {\bibinfo {author} {\bibfnamefont {R.}~\bibnamefont {Dagar}}, \bibinfo {author} {\bibfnamefont {S.}~\bibnamefont {Yadav}}, \bibinfo {author} {\bibfnamefont {M.}~\bibnamefont {Kinha}}, \bibinfo {author} {\bibfnamefont {B.~S.}\ \bibnamefont {Mehra}}, \bibinfo {author} {\bibfnamefont {R.}~\bibnamefont {Rawat}}, \bibinfo {author} {\bibfnamefont {K.}~\bibnamefont {Singh}},\ and\ \bibinfo {author} {\bibfnamefont {D.~S.}\ \bibnamefont {Rana}},\ }\href {https://doi.org/10.1103/PhysRevMaterials.6.074409} {\bibfield  {journal} {\bibinfo  {journal} {Phys. Rev. Mater.}\ }\textbf {\bibinfo {volume} {6}},\ \bibinfo {pages} {074409} (\bibinfo {year} {2022})}\BibitemShut {NoStop}%
\bibitem [{\citenamefont {Solovyev}\ and\ \citenamefont {Kolodiazhnyi}(2016)}]{solovyev_origin_2016}%
  \BibitemOpen
  \bibfield  {author} {\bibinfo {author} {\bibfnamefont {I.~V.}\ \bibnamefont {Solovyev}}\ and\ \bibinfo {author} {\bibfnamefont {T.~V.}\ \bibnamefont {Kolodiazhnyi}},\ }\href {https://doi.org/10.1103/PhysRevB.94.094427} {\bibfield  {journal} {\bibinfo  {journal} {Phys. Rev. B}\ }\textbf {\bibinfo {volume} {94}},\ \bibinfo {pages} {094427} (\bibinfo {year} {2016})}\BibitemShut {NoStop}%
\bibitem [{\citenamefont {May}\ \emph {et~al.}(2020)\citenamefont {May}, \citenamefont {Cao},\ and\ \citenamefont {Calder}}]{May_Mn3Si2Se6_2020}%
  \BibitemOpen
  \bibfield  {author} {\bibinfo {author} {\bibfnamefont {A.~F.}\ \bibnamefont {May}}, \bibinfo {author} {\bibfnamefont {H.}~\bibnamefont {Cao}},\ and\ \bibinfo {author} {\bibfnamefont {S.}~\bibnamefont {Calder}},\ }\href {https://doi.org/https://doi.org/10.1016/j.jmmm.2020.166936} {\bibfield  {journal} {\bibinfo  {journal} {J. Magn. Magn. Mater.}\ }\textbf {\bibinfo {volume} {511}},\ \bibinfo {pages} {166936} (\bibinfo {year} {2020})}\BibitemShut {NoStop}%
\bibitem [{\citenamefont {Cao}\ \emph {et~al.}(2018)\citenamefont {Cao}, \citenamefont {Xu}, \citenamefont {~}, \citenamefont {Yang}, \citenamefont {Zhang}, \citenamefont {Kang}, \citenamefont {He}, \citenamefont {Zheng}, \citenamefont {Liu}, \citenamefont {Wei}, \citenamefont {Li},\ and\ \citenamefont {Cao}}]{Cao_Mn4Ta2O9_2018}%
  \BibitemOpen
  \bibfield  {author} {\bibinfo {author} {\bibfnamefont {Y.}~\bibnamefont {Cao}}, \bibinfo {author} {\bibfnamefont {K.}~\bibnamefont {Xu}}, \bibinfo {author} {\bibfnamefont {Y.}~\bibnamefont {~}}, \bibinfo {author} {\bibfnamefont {W.}~\bibnamefont {Yang}}, \bibinfo {author} {\bibfnamefont {Y.}~\bibnamefont {Zhang}}, \bibinfo {author} {\bibfnamefont {Y.}~\bibnamefont {Kang}}, \bibinfo {author} {\bibfnamefont {X.}~\bibnamefont {He}}, \bibinfo {author} {\bibfnamefont {A.}~\bibnamefont {Zheng}}, \bibinfo {author} {\bibfnamefont {M.}~\bibnamefont {Liu}}, \bibinfo {author} {\bibfnamefont {S.}~\bibnamefont {Wei}}, \bibinfo {author} {\bibfnamefont {Z.}~\bibnamefont {Li}},\ and\ \bibinfo {author} {\bibfnamefont {S.}~\bibnamefont {Cao}},\ }\href {https://doi.org/10.1016/j.jcrysgro.2018.04.007} {\bibfield  {journal} {\bibinfo  {journal} {J. Cryst. Growth}\ }\textbf {\bibinfo {volume} {492}},\ \bibinfo {pages} {35} (\bibinfo {year} {2018})}\BibitemShut {NoStop}%
\bibitem [{\citenamefont {Castellanos~R.}\ \emph {et~al.}(2006)\citenamefont {Castellanos~R.}, \citenamefont {Bern{\`{e}}s},\ and\ \citenamefont {Vega-Gonz{\'{a}}lez}}]{Castellanos_Co4Nb2O9_2006}%
  \BibitemOpen
  \bibfield  {author} {\bibinfo {author} {\bibfnamefont {M.~A.}\ \bibnamefont {Castellanos~R.}}, \bibinfo {author} {\bibfnamefont {S.}~\bibnamefont {Bern{\`{e}}s}},\ and\ \bibinfo {author} {\bibfnamefont {M.}~\bibnamefont {Vega-Gonz{\'{a}}lez}},\ }\href {https://doi.org/10.1107/S1600536806012141} {\bibfield  {journal} {\bibinfo  {journal} {Acta. Crystallogr. E.}\ }\textbf {\bibinfo {volume} {62}},\ \bibinfo {pages} {i117} (\bibinfo {year} {2006})}\BibitemShut {NoStop}%
\bibitem [{\citenamefont {Khanh}\ \emph {et~al.}(2019)\citenamefont {Khanh}, \citenamefont {Abe}, \citenamefont {Matsuura}, \citenamefont {Sagayama}, \citenamefont {Tokunaga},\ and\ \citenamefont {Arima}}]{Khanh_Co4Nb2O9_2019}%
  \BibitemOpen
  \bibfield  {author} {\bibinfo {author} {\bibfnamefont {N.~D.}\ \bibnamefont {Khanh}}, \bibinfo {author} {\bibfnamefont {N.}~\bibnamefont {Abe}}, \bibinfo {author} {\bibfnamefont {K.}~\bibnamefont {Matsuura}}, \bibinfo {author} {\bibfnamefont {H.}~\bibnamefont {Sagayama}}, \bibinfo {author} {\bibfnamefont {Y.}~\bibnamefont {Tokunaga}},\ and\ \bibinfo {author} {\bibfnamefont {T.}~\bibnamefont {Arima}},\ }\href {https://doi.org/10.1063/1.5086894} {\bibfield  {journal} {\bibinfo  {journal} {Appl. Phys. Lett.}\ }\textbf {\bibinfo {volume} {114}},\ \bibinfo {pages} {102905} (\bibinfo {year} {2019})}\BibitemShut {NoStop}%
\bibitem [{\citenamefont {Chang}\ \emph {et~al.}(2023)\citenamefont {Chang}, \citenamefont {Wang}, \citenamefont {Wang}, \citenamefont {Liu}, \citenamefont {You}, \citenamefont {Liu}, \citenamefont {Zheng}, \citenamefont {Shi}, \citenamefont {Lu},\ and\ \citenamefont {Liu}}]{Chang_Co4Nb2O9_2023}%
  \BibitemOpen
  \bibfield  {author} {\bibinfo {author} {\bibfnamefont {Y.}~\bibnamefont {Chang}}, \bibinfo {author} {\bibfnamefont {J.}~\bibnamefont {Wang}}, \bibinfo {author} {\bibfnamefont {W.}~\bibnamefont {Wang}}, \bibinfo {author} {\bibfnamefont {C.}~\bibnamefont {Liu}}, \bibinfo {author} {\bibfnamefont {B.}~\bibnamefont {You}}, \bibinfo {author} {\bibfnamefont {M.}~\bibnamefont {Liu}}, \bibinfo {author} {\bibfnamefont {S.}~\bibnamefont {Zheng}}, \bibinfo {author} {\bibfnamefont {M.}~\bibnamefont {Shi}}, \bibinfo {author} {\bibfnamefont {C.}~\bibnamefont {Lu}},\ and\ \bibinfo {author} {\bibfnamefont {J.-M.}\ \bibnamefont {Liu}},\ }\href {https://doi.org/10.1103/PhysRevB.107.014412} {\bibfield  {journal} {\bibinfo  {journal} {Phys. Rev. B}\ }\textbf {\bibinfo {volume} {107}},\ \bibinfo {pages} {014412} (\bibinfo {year} {2023})}\BibitemShut {NoStop}%
\bibitem [{\citenamefont {Xie}\ \emph {et~al.}(2018)\citenamefont {Xie}, \citenamefont {Zang}, \citenamefont {Ceng}, \citenamefont {Wu},\ and\ \citenamefont {Wang}}]{Xie_Co4Nb2O9_2018}%
  \BibitemOpen
  \bibfield  {author} {\bibinfo {author} {\bibfnamefont {Y.~M.}\ \bibnamefont {Xie}}, \bibinfo {author} {\bibfnamefont {H.}~\bibnamefont {Zang}}, \bibinfo {author} {\bibfnamefont {W.~D.}\ \bibnamefont {Ceng}}, \bibinfo {author} {\bibfnamefont {H.~Y.}\ \bibnamefont {Wu}},\ and\ \bibinfo {author} {\bibfnamefont {C.~C.}\ \bibnamefont {Wang}},\ }\href {https://doi.org/10.1063/1.5039888} {\bibfield  {journal} {\bibinfo  {journal} {Appl. Phys. Lett.}\ }\textbf {\bibinfo {volume} {113}},\ \bibinfo {pages} {082906} (\bibinfo {year} {2018})}\BibitemShut {NoStop}%
\bibitem [{\citenamefont {Xie}\ and\ \citenamefont {Wang}(2022)}]{Xie_Co4Ta2O9_2022}%
  \BibitemOpen
  \bibfield  {author} {\bibinfo {author} {\bibfnamefont {Y.~M.}\ \bibnamefont {Xie}}\ and\ \bibinfo {author} {\bibfnamefont {C.~C.}\ \bibnamefont {Wang}},\ }\href {https://www.sciencedirect.com/science/article/pii/S0925838822004571} {\bibfield  {journal} {\bibinfo  {journal} {J. Alloys Compd.}\ }\textbf {\bibinfo {volume} {906}},\ \bibinfo {pages} {164066} (\bibinfo {year} {2022})}\BibitemShut {NoStop}%
\bibitem [{\citenamefont {Holland}(1963)}]{Holland_thermalconductivity_1963}%
  \BibitemOpen
  \bibfield  {author} {\bibinfo {author} {\bibfnamefont {M.~G.}\ \bibnamefont {Holland}},\ }\href {https://doi.org/10.1103/PhysRev.132.2461} {\bibfield  {journal} {\bibinfo  {journal} {Phys. Rev.}\ }\textbf {\bibinfo {volume} {132}},\ \bibinfo {pages} {2461} (\bibinfo {year} {1963})}\BibitemShut {NoStop}%
\bibitem [{\citenamefont {Carruthers}(1961)}]{Carruthers_thermalconductivity_1961}%
  \BibitemOpen
  \bibfield  {author} {\bibinfo {author} {\bibfnamefont {P.}~\bibnamefont {Carruthers}},\ }\href {https://doi.org/10.1103/RevModPhys.33.92} {\bibfield  {journal} {\bibinfo  {journal} {Rev. Mod. Phys.}\ }\textbf {\bibinfo {volume} {33}},\ \bibinfo {pages} {92} (\bibinfo {year} {1961})}\BibitemShut {NoStop}%
\bibitem [{\citenamefont {Narayanan}\ \emph {et~al.}(2018)\citenamefont {Narayanan}, \citenamefont {Senyshyn}, \citenamefont {Mikhailova}, \citenamefont {Faske}, \citenamefont {Lu}, \citenamefont {Liu}, \citenamefont {Weise}, \citenamefont {Ehrenberg}, \citenamefont {Mole}, \citenamefont {Hutchison}, \citenamefont {Fuess}, \citenamefont {McIntyre}, \citenamefont {Liu},\ and\ \citenamefont {Yu}}]{narayanan_Mn4Ta2O9_2018}%
  \BibitemOpen
  \bibfield  {author} {\bibinfo {author} {\bibfnamefont {N.}~\bibnamefont {Narayanan}}, \bibinfo {author} {\bibfnamefont {A.}~\bibnamefont {Senyshyn}}, \bibinfo {author} {\bibfnamefont {D.}~\bibnamefont {Mikhailova}}, \bibinfo {author} {\bibfnamefont {T.}~\bibnamefont {Faske}}, \bibinfo {author} {\bibfnamefont {T.}~\bibnamefont {Lu}}, \bibinfo {author} {\bibfnamefont {Z.}~\bibnamefont {Liu}}, \bibinfo {author} {\bibfnamefont {B.}~\bibnamefont {Weise}}, \bibinfo {author} {\bibfnamefont {H.}~\bibnamefont {Ehrenberg}}, \bibinfo {author} {\bibfnamefont {R.~A.}\ \bibnamefont {Mole}}, \bibinfo {author} {\bibfnamefont {W.~D.}\ \bibnamefont {Hutchison}}, \bibinfo {author} {\bibfnamefont {H.}~\bibnamefont {Fuess}}, \bibinfo {author} {\bibfnamefont {G.~J.}\ \bibnamefont {McIntyre}}, \bibinfo {author} {\bibfnamefont {Y.}~\bibnamefont {Liu}},\ and\ \bibinfo {author} {\bibfnamefont {D.}~\bibnamefont {Yu}},\ }\href {https://doi.org/10.1103/PhysRevB.98.134438} {\bibfield  {journal} {\bibinfo  {journal} {Phys. Rev. B}\
  }\textbf {\bibinfo {volume} {98}},\ \bibinfo {pages} {134438} (\bibinfo {year} {2018})}\BibitemShut {NoStop}%
\end{thebibliography}

\end{document}